\documentclass[prx, twocolumn, amsmath, amssymb, superscriptaddress, showpacs]{revtex4-1}

\usepackage{amsmath}    
\usepackage{graphicx}   
\usepackage{verbatim}   
\usepackage{color}      
\usepackage[usenames,dvipsnames,svgnames,table]{xcolor}
\usepackage{subfigure}  
\usepackage[hidelinks]{hyperref}   
\usepackage{multirow}
\usepackage{float}
\usepackage{braket}
\usepackage{bbm}
\usepackage{microtype}       
\usepackage{bibunits}
\usepackage{bbold}
\usepackage{booktabs}   

\graphicspath{{images/}}

\newcommand{\puttitle}{Realizing Quantum Convolutional Neural Networks on a Superconducting Quantum Processor to Recognize Quantum Phases}

\begin{document}
		
\begin{abstract}
Quantum computing crucially relies on the ability to efficiently characterize the quantum states output by quantum hardware. Conventional methods which probe these states through direct measurements and classically computed correlations become computationally expensive when increasing the system size. Quantum neural networks tailored to recognize specific features of quantum states by combining unitary operations, measurements and feedforward promise to require fewer measurements and to tolerate errors. Here, we realize a quantum convolutional neural network (QCNN) on a 7-qubit superconducting quantum processor to identify symmetry-protected topological (SPT) phases of a spin model characterized by a non-zero string order parameter. We benchmark the performance of the QCNN based on approximate ground states of a family of cluster-Ising Hamiltonians which we prepare using a hardware-efficient, low-depth state preparation circuit. We find that, despite being composed of finite-fidelity gates itself, the QCNN recognizes the topological phase with higher fidelity than direct measurements of the string order parameter for the prepared~states.
\end{abstract}
		
	\title{\puttitle}
	\author{Johannes Herrmann}
	\email{johannes.herrmann@phys.ethz.ch}
	\affiliation{Department of Physics, ETH Zurich, CH-8093 Zurich, Switzerland}
	\author{Sergi Masot Llima}
	\affiliation{Department of Physics, ETH Zurich, CH-8093 Zurich, Switzerland}
	\author{Ants Remm}
	\affiliation{Department of Physics, ETH Zurich, CH-8093 Zurich, Switzerland}
	\author{Petr Zapletal}
	\affiliation{Department of Physics, Friedrich-Alexander University Erlangen-Nürnberg (FAU), Erlangen, Germany}	
	\author{Nathan A. McMahon}
	\affiliation{Department of Physics, Friedrich-Alexander University Erlangen-Nürnberg (FAU), Erlangen, Germany}	
	\author{Colin Scarato}
	\affiliation{Department of Physics, ETH Zurich, CH-8093 Zurich, Switzerland}
	\author{Fran\c{c}ois Swiadek}
	\affiliation{Department of Physics, ETH Zurich, CH-8093 Zurich, Switzerland}
	\author{Christian Kraglund Andersen}
	\affiliation{Department of Physics, ETH Zurich, CH-8093 Zurich, Switzerland}
	\author{Christoph Hellings}
	\affiliation{Department of Physics, ETH Zurich, CH-8093 Zurich, Switzerland}
	\author{Sebastian Krinner}
	\affiliation{Department of Physics, ETH Zurich, CH-8093 Zurich, Switzerland}
	\author{Nathan Lacroix}
	\affiliation{Department of Physics, ETH Zurich, CH-8093 Zurich, Switzerland}
	\author{Stefania Lazar}
	\affiliation{Department of Physics, ETH Zurich, CH-8093 Zurich, Switzerland}
	\author{Michael Kerschbaum}
	\affiliation{Department of Physics, ETH Zurich, CH-8093 Zurich, Switzerland}
	\author{Dante Colao Zanuz}
	\affiliation{Department of Physics, ETH Zurich, CH-8093 Zurich, Switzerland}
	\author{Graham J. Norris}
	\affiliation{Department of Physics, ETH Zurich, CH-8093 Zurich, Switzerland}
	\author{Michael J. Hartmann}
	\affiliation{Department of Physics, Friedrich-Alexander University Erlangen-Nürnberg (FAU), Erlangen, Germany}	
	\author{Andreas Wallraff}
	\affiliation{Department of Physics, ETH Zurich, CH-8093 Zurich, Switzerland}
	\affiliation{Quantum Center, ETH Zurich, CH-8093 Zurich, Switzerland}
	\author{Christopher Eichler}
    \email{eichlerc@phys.ethz.ch}
	\affiliation{Department of Physics, ETH Zurich, CH-8093 Zurich, Switzerland}
	
	\date{\today}
	
	
	\maketitle
\textbf{Introduction --} 	
Remarkable progress in building quantum hardware~\cite{Arute2019, Zhong2020c, Egan2020, Andersen2020b} has fueled the search for potential applications of both near-term and future error-corrected quantum computers~\cite{Preskill2018, Montanaro2016}, particularly in the simulation of quantum many-body systems~\cite{Feynman1982, Cao2019} and in machine learning~\cite{Biamonte2017, Farhi2018, Cong2019, Beer2020}.
For example, the ability of quantum computers to perform linear algebraic operations more efficiently could provide potential speedups for classical machine learning tasks, such as the ordinary matrix inversion in linear regression models~\cite{Seber2003}. However, dedicated quantum algorithms for this purpose, such as the Harrow, Hassidim and Lloyd~(HHL) algorithm~\cite{Harrow2009}, rely both on executing deep quantum circuits~\cite{Scherer2017} and on loading binary data into a quantum register~\cite{Aaronson2015} to offer practical advantages, which is beyond the reach of currently available quantum hardware.
To load classical data into a quantum register in a more resource-efficient manner and map their features into the high-dimensional Hilbert space to ease classification, quantum circuits parameterized by the input data have been devised and used in quantum support vector machines~\cite{Havlicek2019, Schuld2019} and quantum neural networks~\cite{Jaderberg2021}. However, independent of the specific data embedding scheme, it is still an open question whether tasks aiming at the analysis of classical data can ever fully leverage a quantum computer's capability to process classically unrepresentable amounts of data~\cite{Biamonte2017}.

Promising candidates to harness the capabilities of near-term quantum computers are therefore algorithms which process quantum data directly and for which there is no classical analog~\cite{Biamonte2017}. Quantum computers are beginning to reach a level at which their output states are too complex to be analyzed by classical means~\cite{Arute2019}, suggesting that machine learning techniques which directly process quantum data are expected to become an increasingly important tool to efficiently characterize and benchmark quantum hardware. Examples of specific applications thereof include the principal component analysis of density matrices~\cite{Lloyd2014b}, quantum autoencoders~\cite{Zhang2021f, Bondarenko2020, Romero2017b}, the certification of Hamiltonian dynamics~\cite{Gentile2021, Wiebe2014}, and the detection of entanglement correlations in quantum many-body states~\cite{Cong2019, Farhi2018, Kottmann2021b}.

In this work, we experimentally demonstrate the classification of quantum states with quantum neural networks~\cite{Cong2019} by implementing a quantum algorithm designed to recognize signatures of topological quantum phases~\cite{Pollmann2010, Chen2011c, Pollmann2012}. This challenging task is of great relevance for the study of quantum many-body systems~\cite{Sachdev2001} such as high-temperature superconductors~\cite{Wang2016m}. Previous work in this context has focused on recognizing topological quantum phases from (simulated) measurement data using classical machine learning techniques~\cite{Carrasquilla2017, RodriguezNieva2019, Greplova2020, Lian2019}. Furthermore, topological states have recently been prepared on quantum hardware and analyzed by measuring characteristic observables~\cite{Smith2019, Satzinger2021, Azses2020} such as string order parameters. Here, we experimentally demonstrate a new paradigm to detect symmetry-protected topological states on a 7-qubit quantum device by preparing quantum states within and outside of the SPT phase and by further processing these states with a quantum convolutional neural network to perform quantum phase recognition~\cite{Cong2019}. The QCNN which we implement outperforms the direct measurement of the string order parameter in correctly identifying the topological phase, due to its ability to tolerate both $X$- and $Z$-type Pauli errors while processing weakly perturbed input states.
\\\\
\textbf{Model --} As a model system we consider a family of cluster-Ising Hamiltonians~\cite{Smacchia2011}
	\begin{equation}
		H = -\sum_{i=1}^{N}\left(Z_{i-1}X_{i}Z_{i+1}+h_1 X_i+h_2X_i X_{i+1}\right).
		\label{eq:H}
	\end{equation}
Ground states of \eqref{eq:H} either belong to a topological quantum phase, a paramagnetic phase, or an antiferromagnetic phase depending on the model parameters $\{h_1,h_2\}$. $h_1$ and $h_2$ parametrize the strength of an external field and a nearest-neighbor Ising-type coupling in the model. $\{X_i,Y_i,Z_i\}$ are the Pauli operators acting on the spin at site $i$. We define $Z_0\equiv Z_{N+1}\equiv X_{N+1}\equiv\mathbb{1}$, which models a spin chain with open boundary conditions~\cite{Azses2020}.

\begin{figure}[t]
		\centering
		\includegraphics[width = 0.49\textwidth]{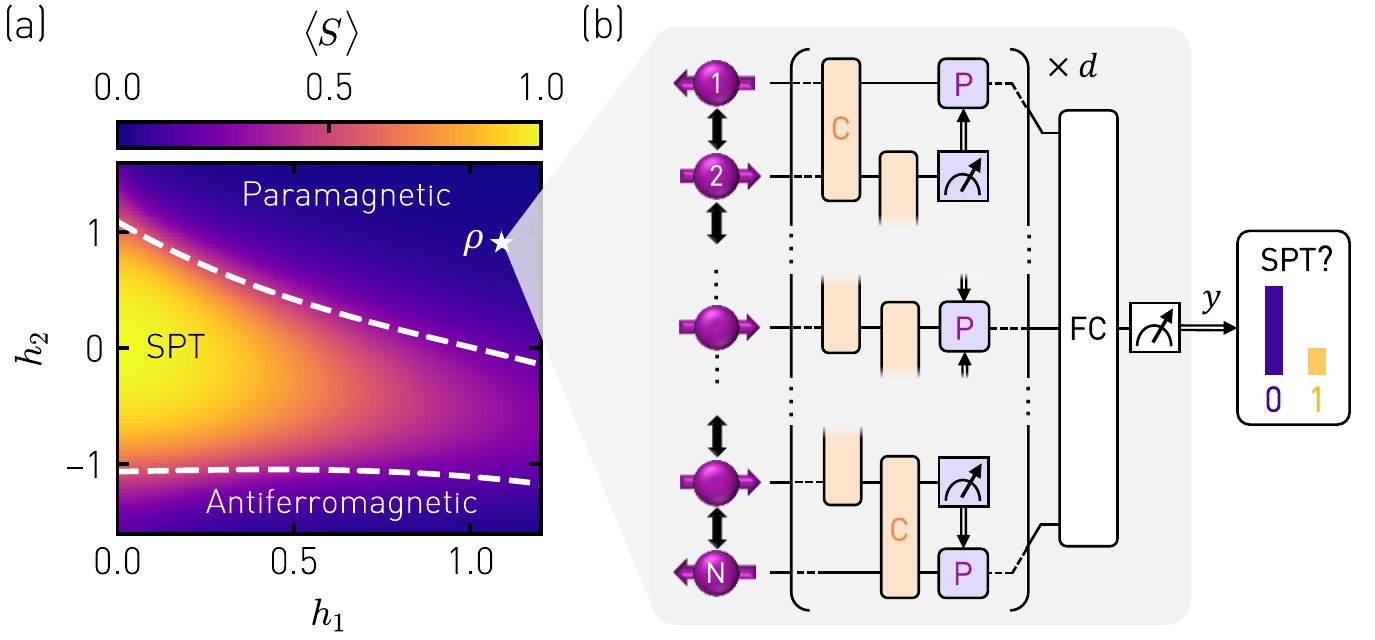}
		\caption{\textbf{Concept of the quantum phase recognition experiment.}~(a) Phase diagram displaying the expectation value of the string order parameter $\braket{\mathcal{S}}=\braket{Z_1X_2X_4X_6Z_7}$ for ground states $\rho$ of a cluster-Ising Hamiltonian~(Eq.~\ref{eq:H}) in the parameter space spanned by $h_1$ and $h_2$ for $N=7$. The white dashed lines indicate the phase boundaries between the symmetry-protected topological~(SPT) phase and the paramagnetic and antiferromagnetic phases, respectively. (b) An unknown state $\rho$ drawn from the phase diagram in (a) is processed by a QCNN to recognize the phase to which it belongs. The QCNN consists of convolutional layers (C) decomposed into two-qubit gates (orange), of pooling layers~(P) implemented as single-qubit operations conditioned on intermediate measurement outcomes (purple), a fully connected  circuit layer~(FC), and the measurement of a single output qubit yielding outcome $y$.}
		\label{fig:qpr}
	\end{figure}

In the thermodynamic limit, the bulk of the Hamiltonian~$H$ commutes with both even $P_\mathrm{e}=\prod_{i}X_{2i}$ and odd $P_\mathrm{o}=\prod_{i}X_{2i+1}$ parity operators, and thus exhibits a $\mathbb{Z}_2 \times\mathbb{Z}_2$ symmetry-protected topological quantum phase~\cite{Verresen2017}, which falls into the same symmetry class as the $S=1$ Haldane phase~\cite{Haldane1983}. The SPT phase is distinguished from the paramagnetic and antiferromagnetic phase by a non-zero expectation value $\braket{\mathcal{S}}$ of the string order parameter~\cite{Pollmann2012}
	\begin{equation}
		  \mathcal{S} = Z_1X_2X_4...X_{N-3}X_{N-1}Z_N.
		 \label{eq:SOP}
	\end{equation}
Corresponding to the experimental situation in this work, we have computed $\braket{\mathcal{S}}$, shown in Fig.~\ref{fig:qpr}(a), using exact diagonalization for a system of $N=7$ spins.
Due to the finite system size, we obtain smooth transitions across the phase boundaries (white dashed lines) determined from the maxima in the second derivative of the energy expectation value $\langle H\rangle$ with respect to $h_2$~\cite{Sachdev2001}.
\\\\
\textbf{Concept of the experiment --}
Conventionally, the phase to which an unknown quantum state $\rho$ belongs is determined by measuring the expectation value of an order parameter $\mathcal{S}$, a process referred to as quantum phase recognition. However, when evaluating the expectation value $\braket{\mathcal{S}}$ by simultaneously, but individually measuring the qubits in their respective basis and by averaging the outcomes over multiple repetitions of the experiment, the sampling complexity increases close to the phase boundaries \cite{Cong2019}. 
Furthermore, under realistic conditions the state $\rho$, which we prepare on the quantum hardware by executing a state-preparation circuit, might be subject to errors, reducing the value of $\langle \mathcal{S}\rangle$.

To overcome the aforementioned limitations, we perform quantum phase recognition by processing the trial states $\rho$ with a QCNN.
The structure of QCNNs, as recently proposed in Ref.~\cite{Cong2019}, is inspired by classical convolutional neural networks widely used e.g. in image or speech recognition.
A generic QCNN consists of alternating convolutional (C) and pooling (P) layers, followed by a fully connected (FC) layer, as schematically shown in Fig.~\ref{fig:qpr}(b). 
The combination of entangling gates applied between neighboring qubits in the convolutional layer, and single-qubit gates conditioned on the outcome of projective measurements in the pooling layer, reduces the number of qubits while retaining characteristic features of the input state vector. After repeating this procedure $d$ times, a unitary operation in the fully connected layer maps the feature of interest onto a single output qubit. In general, QCNNs are parameterized and can be trained to identify specific features of interest.

In our particular case, the QCNN is designed to recognize string order and decide if the input state $\rho$ belongs to the SPT phase or not.  The specific structure of the QCNN is inspired by the multiscale entanglement renormalization ansatz (MERA) representation~\cite{Vidal2008} of the topological cluster state, which is the ground state of $H(h_1=h_2=0)=-\sum Z_{i-1}X_iZ_{i+1}$.
In this case, each pair of convolutional and pooling layers maps a (perturbed) cluster state onto a cluster state of reduced system size, see Appendix~\ref{Theory} for more details. Compared to the originally proposed QCNN~\cite{Cong2019}, we modify the fully connected layer to augment its tolerance to errors and use several gate identities to reduce the quantum gate count, thereby drastically enhancing the performance under NISQ conditions.

	\begin{figure*}[t]
		\centering
		\includegraphics[width = 0.995\textwidth]{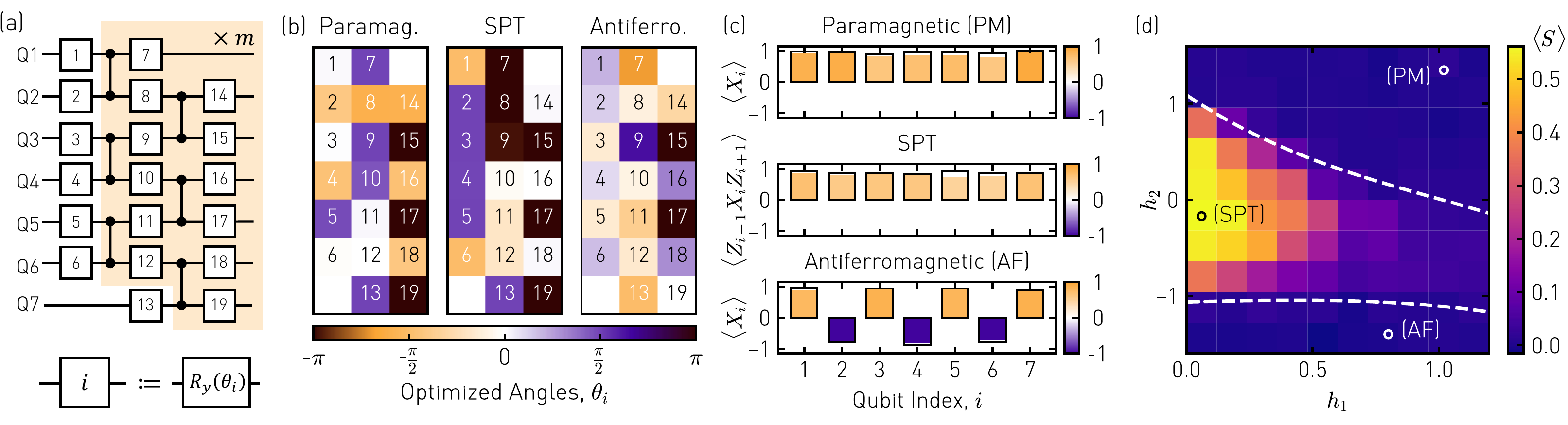}
		\caption{\textbf{Variational ground state preparation.}~(a) Variational quantum circuit parametrized with 19 rotation angles $\pmb{\theta}$ used to prepare approximate ground states of the cluster-Ising Hamiltonian $H$. (b) Rotation angles $\theta_i$ found by an optimization algorithm on a conventional computer for three example states~$\{h_1,h_2\}$ in the paramagnetic (PM)~$\{1.1,1.4\}$, SPT~$\{0.0,-0.2\}$, and antiferromagnetic (AF)~$\{0.8,-1.4\}$ phase. (c)~Measurement~(solid bars) of the indicated expectation values along the qubit array in comparison to the simulated values~(wire frames) for the three states prepared using the rotation angles in~(b). (d)~Measured string order parameters~$\braket{\mathcal{S}}$ for all prepared variational states vs. Hamiltonian parameters $h_1$ and $h_2$. Open circles indicate the three example states presented specifically in~(b)~and~(c).}
		\label{fig:var_gnd_prep}
	\end{figure*}

For our experimental study carried out on a 7-qubit device, we combine two complementary elements. First, we prepare approximate ground states of the cluster-Ising Hamiltonian $H$ by executing variational state preparation circuits. Second, we use those states as an input to the QCNN to demonstrate its capability to recognize the SPT phase and compare it to directly measuring $\langle S\rangle$.
\\\\
\textbf{Variational state preparation --}
To prepare approximate ground states of $H$ for the entire parameter range $\{h_1,h_2\}$ displayed in Fig.~\ref{fig:qpr}(a), we use a low-depth, variational state preparation circuit $U(\pmb{\theta})$ composed of three layers of single-qubit rotations $R_y(\theta_i)$ parametrized by 19 independent rotation angles $\theta_i$ and two layers of conditional-Z (CZ) gates interleaved with the single-qubit gates, see Fig~\ref{fig:var_gnd_prep}(a)~\cite{BravoPrieto2020}. We implement both types of gates directly on the quantum hardware, see Appendix~\ref{Methods} for details.
Compared to previous experiments on superconducting quantum hardware, in which an exact matrix product state representation~\cite{Smith2019} was used to prepare ground states of $H$ for specific combinations $\{h_1,h_2\}$,
the use of a variational circuit allows us to prepare approximate ground states for any parameter set~$\{h_1,h_2\}$.	

To determine the variational parameters~$\pmb{\theta}=\{\theta_1,...,\theta_{19}\}$ corresponding to an approximate ground state of a specific $H(h_1,h_2)$, we minimize the energy expectation value~$\braket{H}$ in a conventional computer simulation with respect to the simulated output state $\ket{\pmb{\theta}}=U(\pmb{\theta})\ket{0}$ by using a gradient based L-BFGS optimizer~\cite{Byrd1995}.
As an acceptance criterion for the convergence of the optimization algorithm, we compute the fidelity $F=\left|\braket{\psi_0|\pmb{\theta}_\mathrm{opt}}\right|^2$ of the variational state~$\ket{{\pmb{\theta}}_\mathrm{opt}}$ with respect to the exact ground state~$\ket{\psi_0}$, which, for $N=7$, can be found using exact diagonalization. We repeat the optimization procedure with different initial values until $F$ exceeds 90$\,\%$ solely being limited by the finite variational circuit depth of $m=1$, see Appendix~\ref{Methods}. To make the state preparation circuit less susceptible to $T_1$ errors, we then compute an equivalent set of rotational angles $\tilde{\pmb{\theta}}_{\rm opt}$ yielding the same $U(\tilde{\pmb{\theta}}_{\rm opt})=U({\pmb{\theta}}_{\rm opt})$~\cite{Fontana2020a}, but keeping the individual qubits preferentially in their respective ground state in the beginning of the state preparation sequence, see Appendix~\ref{Methods} for details. This procedure avoids rotation angles close to $\pi$ in the first layer of single qubit rotations, which becomes apparent in the three examples shown in Fig.~\ref{fig:var_gnd_prep}(b) by the absence of large rotation angles in the first column.  The example state from the paramagnetic phase (PM) features rotation angles summing to approximately~$\pm\pi/2$ for each qubit individually. For the example state from the SPT phase, all qubits are initially rotated by an angle close to $\pm\pi/2$, which, together with the subsequent layers of entangling CZ gates, results in an approximate cluster state~\cite{Azses2020}.

For the rotation angles~$\tilde{\pmb{{\theta}}}_{\mathrm{opt}}$ found in computer simulation, we execute the corresponding state preparation circuits on a 7-qubit superconducting quantum device featuring individual control and readout of all qubits, see Appendix~\ref{Setup} for details.  We realize single-qubit rotations by applying microwave pulses of controlled amplitude and phase and implement two-qubit CZ gates with flux pulses bringing the state~$\ket{11}$ into resonance with the non-computational state $\ket{20}$~\cite{Strauch2003, DiCarlo2009}, where $\ket{n_{i},n_{j}}$ denote the states of the involved qubits in the Fock basis.
To assure that all qubits are in their respective ground state at the beginning of each sequence, we perform a preselection readout and reject those measurement runs in which we found at least one of the qubits to be in the excited state, resulting in an overall acceptance probability of~$\sim91\,\%$. We perform measurements in the $X$~basis by prepending an $R_y(\pi/2)$ rotation to the  respective qubits before performing standard dispersive readout in the $Z$~basis. To mitigate the effect of readout infidelities on the order of 1.6\,\% per qubit in averaged measurement observables, we multiply the vector of probabilities $p_i$ of sampling the bitstring $x_i$ by the inverse of the assignment probability matrix $M$ to obtain $\tilde{\pmb{p}}= M^{-1}{\pmb{p}}$, from which we evaluate expectation values, see Appendix~\ref{ControlReadout} for details. 

To characterize our state preparation, we measure a set of local expectation values and compare the results with those obtained from Kraus operator simulations, which take qubit dissipation and dephasing, as well as measured CZ gate errors and nearest-neighbor residual ZZ coupling into account, see Fig.~\ref{fig:var_gnd_prep}(c). For states in the paramagnetic (antiferromagnetic) phase, we find, as expected, a non-zero magnetization $\braket{X_i}$ along the entire spin chain with equal (alternating) sign. The key signature of states in the SPT phase is the non-zero expectation value of Pauli strings~$\braket{Z_{i-1}X_iZ_{i+1}}$. In all three cases, the average deviation between measured and simulated expectation values is below 5.8\,\% and can likely be attributed to small additional coherent control errors, which we do not account for in the Kraus operator simulation.

\begin{figure}[t]
	\centering
	\includegraphics[width = 0.49\textwidth]{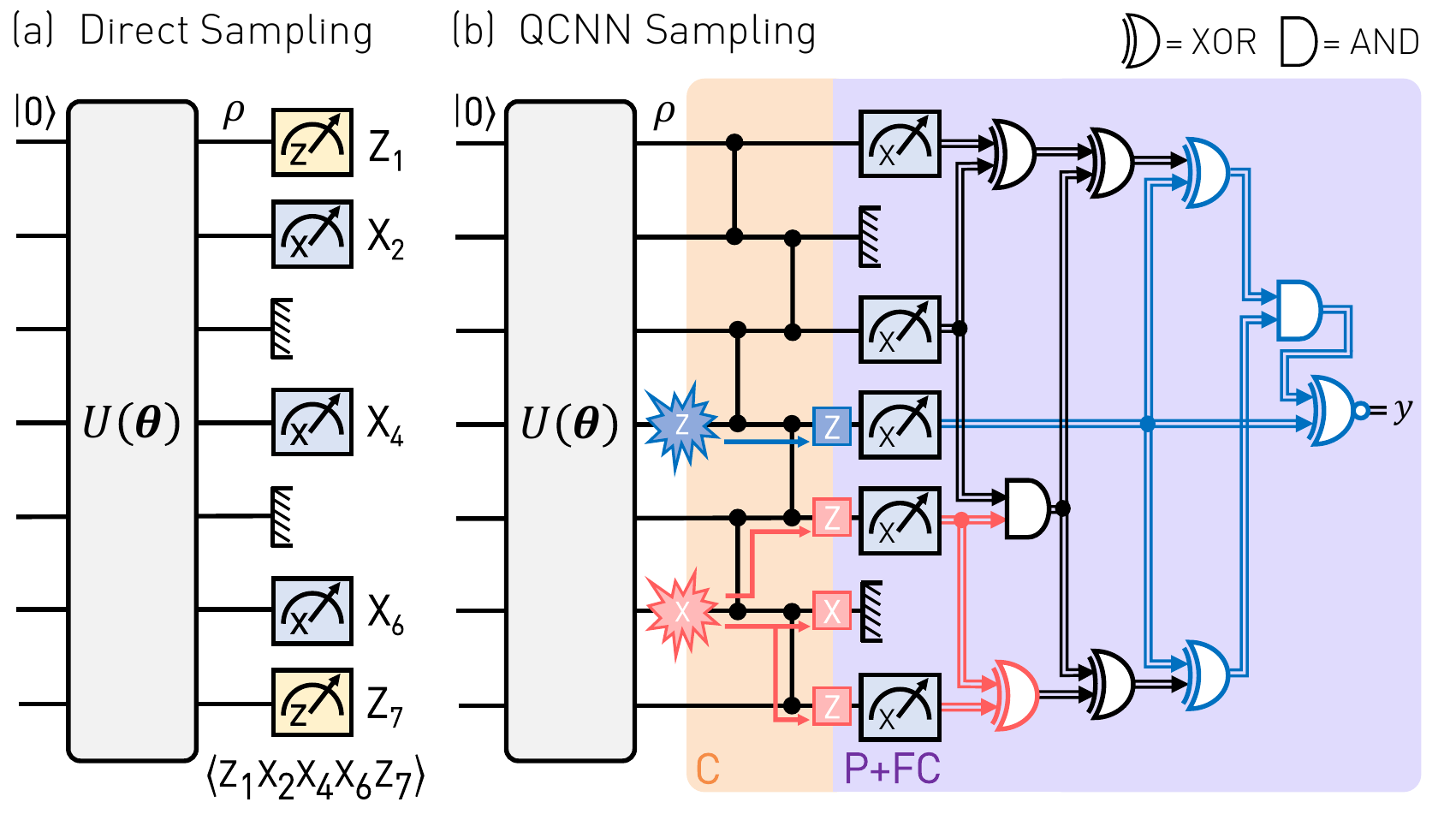}
	\caption{\textbf{Quantum phase recognition circuits}. (a) Quantum circuit for the case in which the qubits are measured in the indicated basis, directly after executing the state preparation circuit $U({\pmb{\theta}})$, to evaluate $\mathcal{S}=Z_1X_2X_4X_6Z_7$. (b)~QCNN circuit consisting of a convolutional layer~(C) of CZ gates (orange), and a pooling~(P) and fully connected~(FC) layer implemented as a measurement in the $X$ basis with outcome $\pmb{x}$, followed by a Boolean function $f(\pmb{x})$, here, represented by a logic circuit expressed in terms of AND and XOR gates~(purple). An exemplary~$X$~($Z$) error occurring on qubit six~(four) and its propagation through the QCNN is highlighted in red (blue).}
	\label{fig:qcnn_circ}
\end{figure}

To map out the full quantum phase diagram shown in Fig.~\ref{fig:var_gnd_prep}(d), we proceed with measuring the string order parameter~$\braket{\mathcal{S}}$ for 10$\times$10 combinations of $\{h_1,h_2\}$ by directly sampling from the output of the state preparation circuit as indicated in Fig.~\ref{fig:qcnn_circ}(a). We find the measured phase diagram to show all qualitative features of the exact one shown in Fig.~\ref{fig:qpr}(a). The reduction in the overall contrast, which becomes apparent in the different amplitude scaling of $\langle\mathcal{S}\rangle$, stems from errors due to decoherence and two-qubit gate errors as confirmed by Kraus operator simulations, see Appendix~\ref{Methods} for details.
Most importantly, some of those errors can be tolerated when inferring the quantum phase of a prior unknown state by applying a QCNN algorithm rather than measuring~$\braket{\mathcal{S}}$ directly, as we will show in the following.
\\\\	
\textbf{Quantum phase recognition --}
Instead of measuring the expectations~$\braket{\mathcal{S}}$ directly after state preparation, we now employ the QCNN to detect the SPT phase. For this purpose, we process the prepared quantum states by the QCNN depicted in Fig.~\ref{fig:qcnn_circ}(b) and evaluate the expectation value $2\langle y\rangle - 1$ of the single output bit~$y$, which corresponds to measuring the expectation value of a multi-scale string order parameter ${\mathcal{S}_{\rm M}}$, consisting of a weighted sum of Pauli strings, see Eq.~(\ref{eq:MSOP7}) in Appendix~\ref{Theory} for an explicit expression. Measuring $\mathcal{S}_{\rm M}$ instead of $\mathcal{S}$ is advantageous because weakly perturbed cluster states still yield $\mathcal{S}_{\rm M}=+1$ while the same states have a finite probability to yield $\mathcal{S}=-1$ which reduces the fidelity of the respective expectation value. Using the QCNN allows us to measure all the individual strings in $\mathcal{S}_{\rm M}$ simultaneously, thereby reducing the number of required measurements by an amount which scales double-exponentially with the QCNN circuit depth $d$ and is equal to~3 for the case $d=1$, see Appendix~\ref{Theory} for details.

\begin{figure*}[t]
	\centering
	\includegraphics[width = 0.99\textwidth]{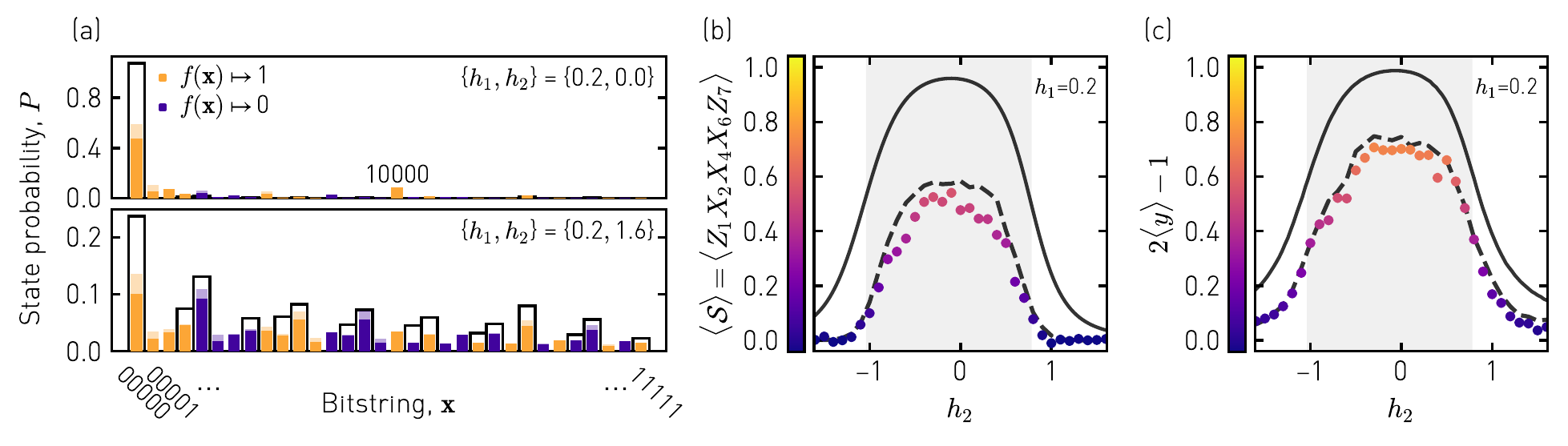}
	\caption{\textbf{Performance of the QCNN.} (a) Probability to sample bitstrings $\pmb{x}$ after having applied the convolutional layer~(compare Fig.~3(b)) for the two indicated Hamiltonian parameter sets $\{h_1,h_2\}$. Bitstrings mapped onto 1 (0) by the function $f(\pmb{x})$ are colored in orange (purple) and the expected probabilities from a Kraus operator simulation in the corresponding light color, whereas the ideal values are depicted as black wire frames. (b)~Expectation value $\braket{\mathcal{S}}=\braket{Z_1 X_2X_4X_6Z_7}$ measured directly after variational state preparation (compare Fig.~3(a)) vs. $h_2$ for fixed $h_1=0.2$, in comparison to the ideal values (solid line) and simulated values taking decoherence into account (dashed line). The SPT phase is indicated in light gray. (c)~Expectation value $2\braket{y}-1$ measured after applying the QCNN for the same parameters as in (b) compared to values extracted from a Kraus operator simulation of the QCNN circuit (dashed line) and the ideal value (solid line).}
	\label{fig:qcnn_data}
\end{figure*}

We construct the QCNN by making two modifications to the quantum circuit proposed in Ref.~\cite{Cong2019}. First, we perform the pooling and fully-connected layers as a Boolean function $f(\pmb{x})$ after having performed a projective measurement, which greatly reduces the total quantum gate count. Second, we extend the fully-connected layer to map the measured bitstring $\pmb{x}=(x_1,x_3,x_4,x_5,x_7)$ onto a single output bit $y$, such that it not only tolerates $X$ errors, but also $Z$ errors, provided the errors are sufficiently sparse. For example, an $X$ error occurring on qubit six and a $Z$ error on qubit four prior to the convolutional layer invert bits $x_5, x_7$ and $x_4$, which is corrected by the function $f(\pmb{x})$, see red and blue colored paths in Fig.~\ref{fig:qcnn_circ}(b).
	
To investigate the QCNN's error-correcting capability in more detail, we sample $\pmb{x}$ after having performed the convolutional layer for two different ground states chosen from the SPT and paramagnetic phase, respectively, and obtain the probability distributions shown in Fig.~\ref{fig:qcnn_data}(a). For the state in the SPT phase (top panel), we find a high probability of 0.47 to sample $(00000)$, which is expected because the ideal cluster state
corresponding to the ground state of $H(h_1=h_2=0)$
is mapped onto~$|$+++++$\rangle$ by the disentangling CZ gates of the quantum convolutional layer. However, due to the non-zero value of $h_1=0.2$ and the presence of noise in the quantum circuit, we also measure other bitstrings with non-zero probability, most notably $(10000)$. Most importantly, a large fraction of those bitstrings is correctly mapped onto $y=1$ by the function $f$, thereby counteracting a quantum phase misclassification in those cases.
For the paramagnetic example (bottom panel), we find the sampled bitstrings to be more uniformly distributed and, correspondingly, $y$ to result equally likely in~0 or~1. In both examples, we find the measured probability distributions to be in good agreement with the simulated ones, taking decoherence, two-qubit gate imperfections and readout errors into account.

We finally quantify the performance of the QCNN in correctly identifying the SPT phase by comparing the value $2\langle y\rangle - 1$ obtained from the QCNN to the value~$\braket{\mathcal{S}}$ obtained from direct sampling, see Figs.~\ref{fig:qcnn_data}(b) and~(c).	
In particular, we determine both quantities across the phase boundaries separating the SPT phase~(light gray) from the paramagnetic and antiferromagnetic phases, respectively, by varying $h_2$ for constant $h_1=0.2$. In both cases the measured values (dots) approach zero for the paramagnetic and antiferromagnetic phases and take a non-zero value reaching 0.70 in the SPT phase. Compared to the ideal values (solid lines) and as a result of error events, the overall fidelity is reduced -- an effect which is well-explained by Kraus operator simulations of the respective quantum circuits (dashed lines), which also identify two-qubit gate imperfections as the most prominent error contributor, see Appendix~\ref{Methods} for details on the simulations. Most importantly, we find the QCNN output $2\langle y\rangle - 1$ to generally have a higher fidelity compared to the directly measured value $\braket{\mathcal{S}}$ and to be much closer to the ideal case. This enhancement of performance provides clear evidence for the robustness of the QCNN against errors.
\\\\	
\textbf{Discussion and outlook --}
By implementing a QCNN on a superconducting quantum processor, we have demonstrated its capability to efficiently recognize quantum phases. With further advances in qubit number and circuit depth, we expect QCNNs to become an important diagnostic tool to characterize output states of NISQ devices, which are increasingly challenging to analyze with classical computing.
Such applications will benefit from the predicted increased sampling efficiency at phase boundaries, which should become accessible with increased systems size $N$~\cite{Cong2019}. This scaling advantage can be understood by expressing the output of a QCNN by an equivalent weighted sum of string order parameters, the number of which scales exponentially with $N$. The QCNN thus allows one to simultaneously measure the sum of all those terms. An interesting direction to be explored in future work includes the trainability of parameterized QCNNs. This also becomes relevant in the context of using QCNNs to learn optimal strategies for quantum error correction.
\\\\	
\textbf{Acknowledgments --}
We thank Mihai Gabureac for contributions to the device fabrication, Arne Wulff for technical contributions to the measurement setup, and Frank Pollmann, Stefan Filipp, Federico Roy and Michele Collodo for input on the manuscript. The authors acknowledge financial support by the EU program H2020-FETOPEN project 828826 Quromorphic, by the EU Flagship on Quantum Technology H2020-FETFLAG2018-03 project 820363 OpenSuperQ, by the Office of the Director of National Intelligence (ODNI), Intelligence Advanced Research Projects Activity (IARPA), via the U.S. Army Research Office grant W911NF-16-1-0071, by the National Centre of Competence in Research Quantum Science and Technology (NCCR QSIT), a research instrument of the Swiss National Science Foundation (SNSF), by the SNFS R’equip grant 206021-170731 and by ETH Zurich. The views and conclusions contained herein are those of the authors and should not be interpreted as necessarily representing the official policies or endorsements, either expressed or implied, of the ODNI, IARPA, or the U.S. Government.
\\\\
\textbf{Author contributions --}
J.H. and C.E. conceptualized the experiment. J.H. and S.M.L. carried out the experiment and analyzed the data. N.A.M., P.Z. and M.J.H. developed the implemented QCNN and provided theory input. P.Z. and J.H. performed the numerical simulations. J.H., F.S., A.R. and C.K.A. designed the device. G.J.N., D.C.Z. and S.K. contributed to the device fabrication. C.S., C.H., A.R., N.L., S.L. and M.K. contributed to the experimental control software and the device tuneup. M.J.H., A.W. and C.E. supervised the work. J.H. and C.E. wrote the manuscript with input from all co-authors.
\\\\
\textbf{Competing interests --}
The authors declare no competing interests.
\\\\
\section*{Supplementary Information}
\section{Experimental Setup}
\label{Setup}	
\begin{figure}[b]
	\centering
	\includegraphics[width = 0.46\textwidth]{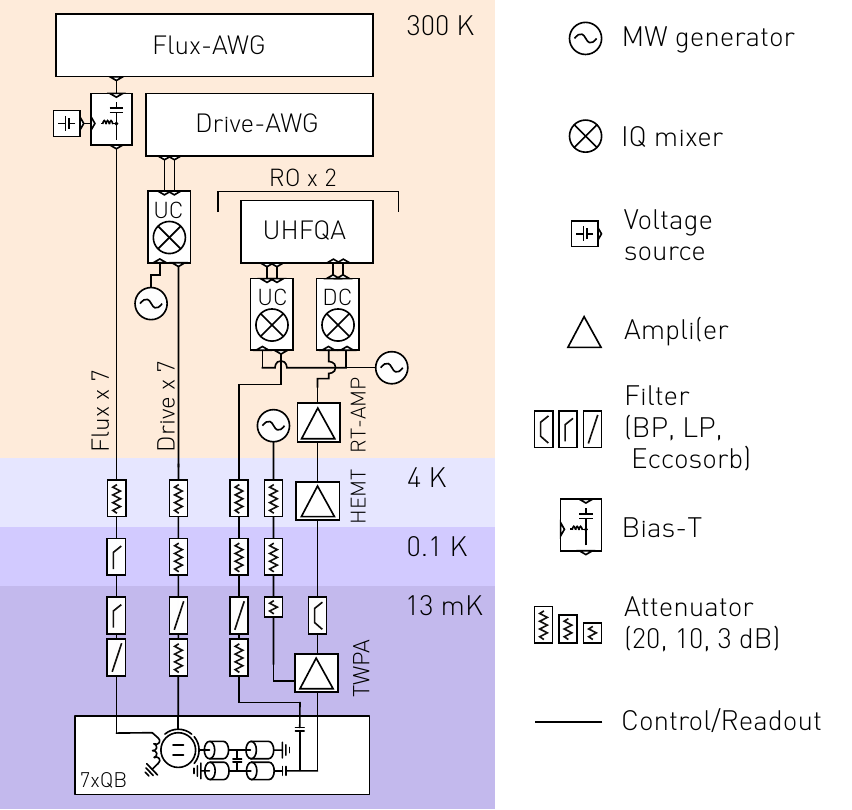}
	\caption{Schematic of the control electronics and wiring setup. For details see main text.}
	\label{fig:setup}
\end{figure}

\begin{figure*}[t]
	\centering
	\includegraphics[width = 0.97\textwidth]{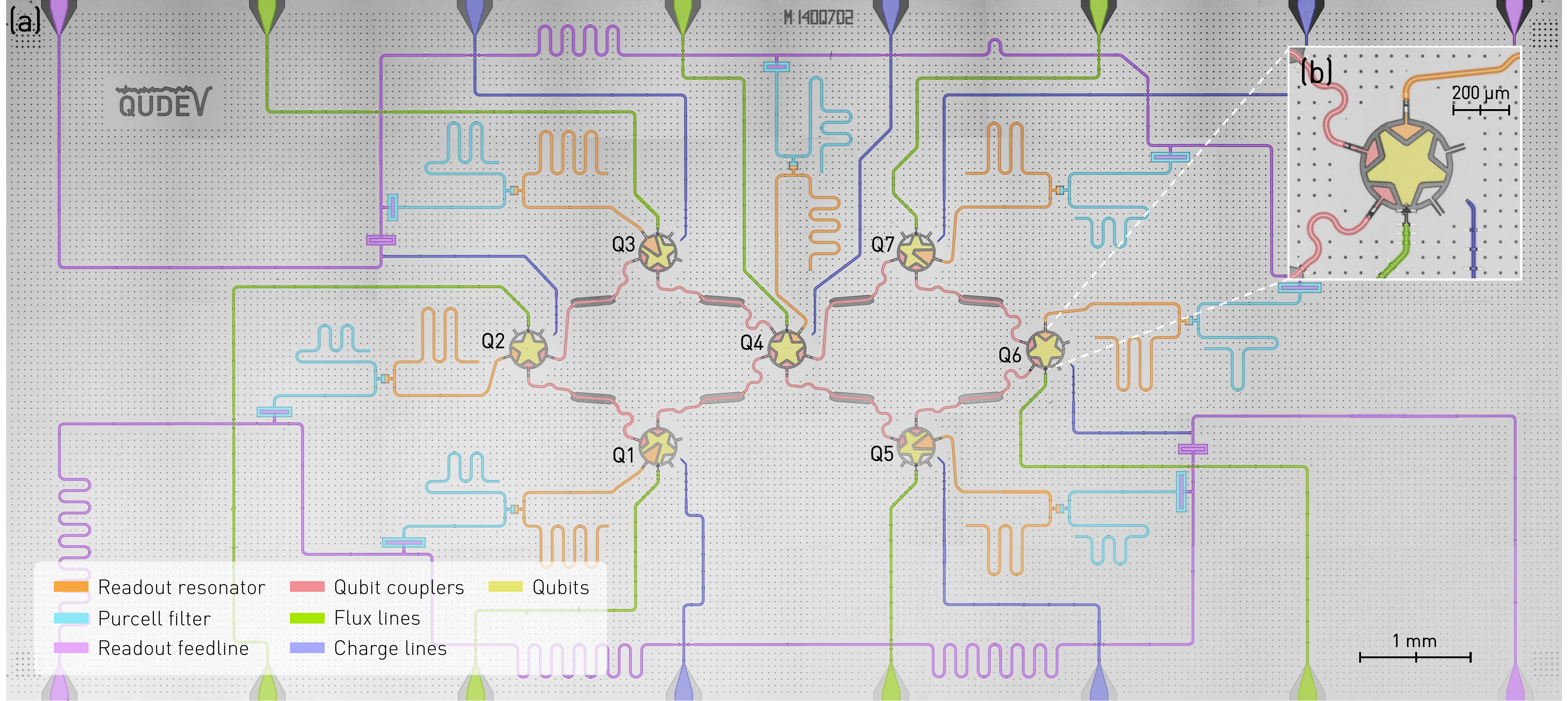}
	\caption{(a) False-color micrograph of the 7-qubit quantum processor with individual elements specified in the legend on the bottom left.  (b)~Enlarged view of the transmon qubit~Q6 and its connecting lines.}
	\label{fig:chip_false_color}
\end{figure*}

\textbf{Device fabrication --} We fabricated the 7-qubit quantum processor, shown in Fig.~\ref{fig:chip_false_color}, by sputtering a Niobium thin film onto a high-resistivity intrinsic Silicon substrate in a process similar to the one described in Ref.~\cite{Andersen2020b}. After patterning the Niobium base layer using photolithography and reactive-ion etching, we fabricate airbridges to establish well-connected ground planes and to enable crossings of signal lines. We fabricate Josephson junctions by shadow evaporation of aluminum through a resist mask defined by electron-beam lithography.
\\\\
\textbf{Device Parameters --} We extract the qubit and readout circuit parameters, summarized in Table~\ref{tab:qb_paras}, using standard spectroscopy and time-domain measurements.
We extract the quantum measurement efficiencies using the methods described in Refs.~\cite{Bultink2018, Heinsoo2018}.	
\\\\
\textbf{Wiring and instrumentation --} We install the device at the base plate of a cryogenic measurement setup (13\,mK) and connect it to the control and measurement electronics as shown in Fig.~\ref{fig:setup}. We control the individual qubit frequencies by threading a magnetic flux through the superconducting quantum interference device (SQUID) loop via a current applied to an inductively coupled flux control line. 
The flux control signal is composed of a constant offset superimposed with pulses controlled on the nanosecond timescale. The constant offset, which we generate using a voltage source (SRS SIM928) and a 1\,k$\Omega$ bias resistor in series, tunes the qubit to its idle frequency. Fast pulses, which we generate using an arbitrary waveform generator (Tektronix AWG5014C), are used to activate two-qubit gates (see Appendix~\ref{ControlReadout} for details). We combine both signals at room-temperature using a bias-tee (Mini-Circuits ZFBT-4R2GW+) with a primary timeconstant of $\sim$18\,$\mu$s to filter out low-frequency noise present at the output of the AWG. 
We achieve XY-control of all seven qubits by up-converting the in-phase and quadrature components of an intermediate frequency signal with analog IQ-mixer modules~(Zurich Instruments HDIQ). Qubit drive pulses are provided by two 8-channel AWGs with a sample rate of 2.4\,GSa/s~(Zurich Instruments HDAWG).

\begin{table*}[t]
	\begin{tabular}{l c c c c c c c}
		\toprule
		& Q1 & Q2 & Q3 & Q4 & Q5 & Q6 & Q7 \\ \midrule
		Qubit idle frequency, $\omega_\mathrm{q}/2\pi$ (GHz)  &4.183&5.949&4.453 &5.881 &4.522 & 6.071&4.200\\
		Qubit anharmonicity, $\alpha_\mathrm{q}/2\pi$ (MHz) &-181 &-170 &-177 &-173 &-178 &-166 &-179 \\
		Lifetime, $T_1$($\mu$s) & 38.1 & 13.5 & 19.8 & 13.7 & 19.8 & 16.4 & 15.6\\
		Ramsey decay time, $T_2^*$($\mu$s) & 20.7 & 12.9 & 14.3  & 10.1 & 13.1& 19.1&8.2\\
		Echo decay time, $T_2^\mathrm{e}$($\mu$s) & 33.2 & 14.1 & 16.7 & 10.4&18.5 & 22.1& 20.5\\
		Readout resonator frequency, $\omega_\mathrm{r}$(GHz) & 6.668 & 7.089 & 6.603 & 7.213 & 6.904 & 6.994 & 6.812\\
		Readout linewidth, $\kappa_\mathrm{eff}$(MHz) & 11 & 7 & 8 & 12& 10& 11 & 9 \\
		Dispersive Shift, $\chi/2\pi$(MHz) &  -3.5 & -3.6 & -3.0  &-1.9 &-2.0 & -2.4& -2.5\\
		Thermal population, $P_\mathrm{th} (\%)$ & 3.1 & 0.7 &  2.7 & 0.7&1.3 & 1.3&1.6\\
		Individual readout assignment prob. (\%) & 99.2 & 98.7 & 98.5  & 97.7& 99.4 &99.2 &98.4\\
		Multiplexed readout assignment prob. (\%) & 99.1 & 98.2 & 98.2  & 97.4& 96.4 &97.7 &98.2\\
		Measurement efficiency, $\eta$ (\%) & 34.0 & 31.3 & 15.4  & 15.9& 23.4 & 25.3&14.7\\
		\bottomrule
	\end{tabular}
	\caption{Measured parameters of the seven qubits.}
	\label{tab:qb_paras}
\end{table*}

We perform frequency-multiplexed qubit readout using two FPGA-based control systems with a sampling rate of 1.8\,GSa/s~({Zurich Instruments UHFQA}). The multi-chromatic readout pulse is upconverted to the readout resonator frequency band and routed through a highly attenuated RF line to the readout line on the chip. At the output, each one of the two frequency-multiplexed readout lines is connected to an amplification chain consisting  of a traveling wave parametric amplifier~(TWPA)~\cite{Macklin2015} at base, a cryogenic high-electron mobility transistor~(HEMT) at 4\,K, and additional low-noise amplifiers at room temperature~(RT-AMP). The amplified signals are finally downconverted to an intermediate frequency band, digitized and integrated by the weighted integration units of the~{UHFQA}s.
\\\\

\section{Control and Readout}
\label{ControlReadout}
\begin{figure*}[t]
	\centering
	\includegraphics[width = 0.85\textwidth]{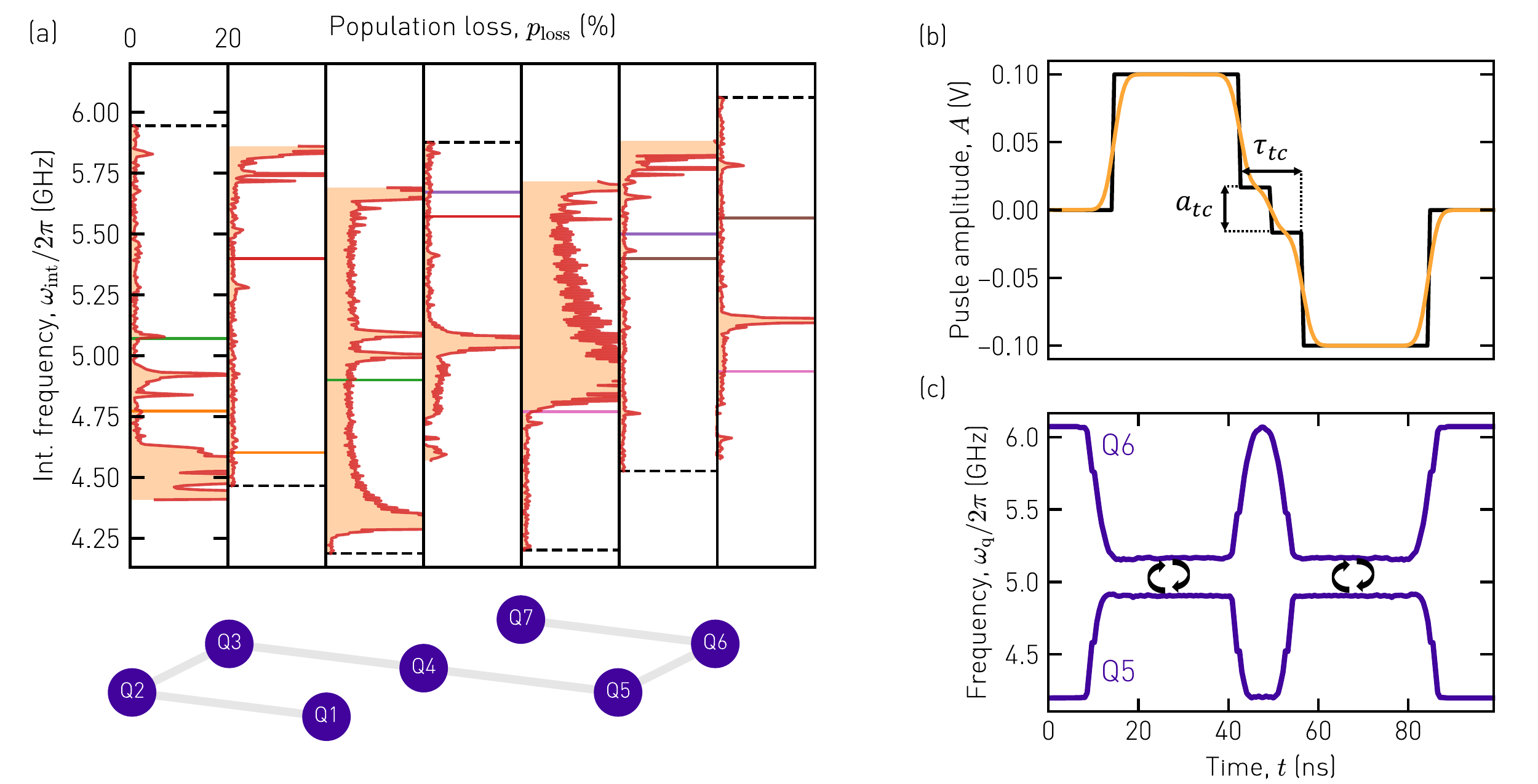}
	\caption{\textbf{Two-qubit gate implementation.} (a) Measured population loss for all qubits when being tuned from the idle frequency (dashed horizontal line) to the frequency $\omega_\mathrm{int}$ for a duration of 100\,ns. Colored lines indicate the chosen interaction frequencies during two-qubit gates for the respective qubit pair. (b)~Parametrization of the net-zero transition-controlled flux pulse, with an adjustable transition amplitude $a_{tc}$ in the transition part of length $\tau_{tc}$ between the two halves of the net-zero pulse before (black) and after (orange) applying an additional Gaussian filtering with $\sigma=0.5$\,ns. (c)~Measured time-dependent frequencies of qubits Q5 and Q6 in response to the flux pulses shown in~(a) with predistortion applied. Frequencies at which the $\ket{11}$ and $\ket{20}$ states interact resonantly are indicated by round arrows.}
	\label{fig:2qubit_gates}
\end{figure*}
\begin{figure*}[t]
	\centering
	\includegraphics[width = 0.85\textwidth]{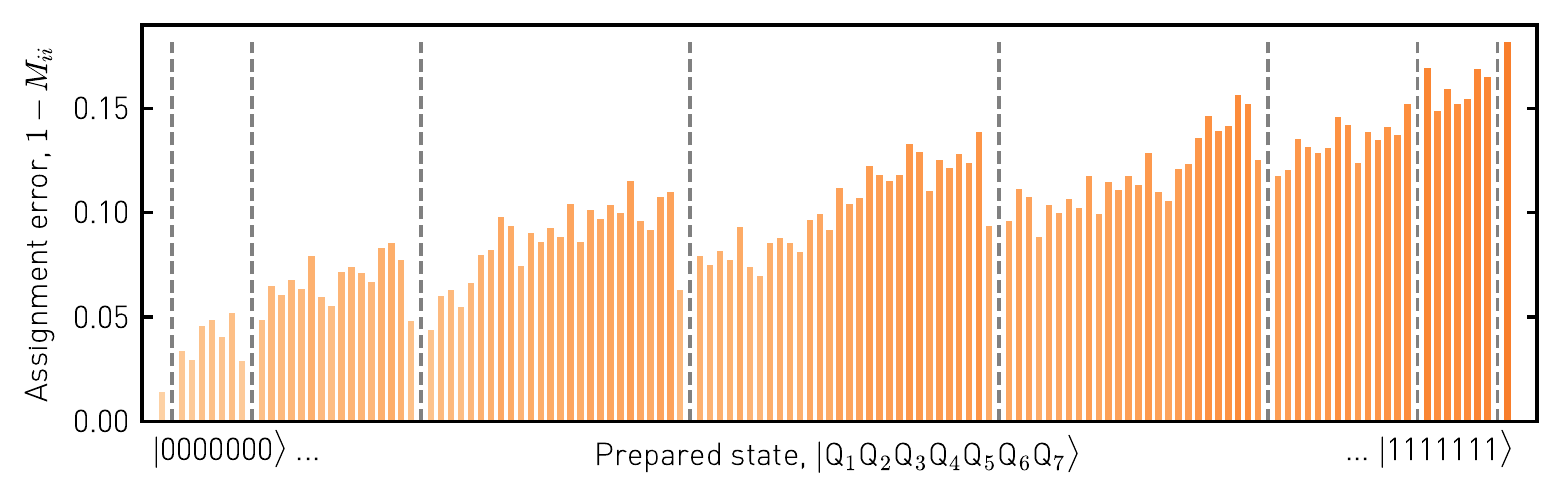}
	\caption{\textbf{Multiplexed qubit readout.} Assignment error $1-M_{ii}$, for the simultaneous single-shot readout of all seven qubits. The prepared states are sorted by the number of qubit excitations (separation by the dashed gray lines) and arranged in binary ascending order within their subdivision.}
	\label{fig:ssro}
\end{figure*}
\textbf{Single-qubit gates -- }
To achieve XY-control we generate pulses having a carrier frequency resonant with the qubit transition frequency and an envelope following a DRAG pulse parametrization to reduce leakage into non-computational states~\cite{Motzoi2009}. We choose a pulse width of $\sigma=10$\,ns and truncate to a pulse length of $\tau=5\sigma=50$\,ns. To implement rotations $R_y(\theta_i)$ with a continuously adjustable target angle $\theta_i \in ]-\pi, +\pi]$, we linearly scale the calibrated $\pi$-pulse amplitude $A_\pi$ to $A_\theta =  A_\pi \theta/\pi$.
\\\\
\textbf{Two-qubit gates --}
We perform two-qubit controlled-Z (CZ) gates by harnessing the in-situ tuneability of the transition frequencies of the qubits.
Our choice of idle frequencies places the qubits in two main frequency bands, where neighboring qubits alternate between upper and lower sweet spot, see dashed lines in Fig.~\ref{fig:2qubit_gates}(a). The large detuning of those frequency bands by approximately 1.7\,GHz keeps residual ZZ coupling during idle times~\cite{Collodo2020, Krinner2020a} below 15\,kHz for all qubit pairs, see Appendix~\ref{GatePerformance} for details. In addition, biasing the qubits to one of their sweet spots reduces the susceptibility to flux noise.
To activate a two-qubit CZ gate, we tune both participating qubits to an intermediate interaction frequency such that the $\ket{11}$ and the non-computational $\ket{20}$ state become resonant~\cite{DiCarlo2009, Strauch2003}. By choosing an intermediate interaction frequency, we avoid frequency collisions with neighboring qubits, which are not supposed to participate in the gate.
The exact choice of interaction frequencies, see colored lines in Fig.~\ref{fig:2qubit_gates}(a), is governed by the avoidance of frequency-dependent population loss, which is most likely caused by the interaction with two-level systems residing at the material interfaces and inside the tunnel junctions~\cite{Lisenfeld2019, Klimov2018}. We characterize the population loss for each qubit individually, by measuring the remaining population of an initially prepared excited state after a 100\,ns long rectangular flux pulse smoothed with a Gaussian filter of width $\sigma=0.5$\,ns, which tunes the respective qubit to the variable interaction frequency $\omega_\mathrm{int}$, see Fig.~\ref{fig:2qubit_gates}(a).

The flux pulses activating the two-qubit gate interaction have a net-zero transition-controlled~(NZTC) pulse shape, as depicted in Fig.~\ref{fig:2qubit_gates}(b), consisting of a NZ pulse to assure flux pulse repeatability and an additional adjustable amplitude step to provide control over the transition part~\cite{Rol2019, Negirneac2021}.
We parametrize the transition part of the pulse with an individual length~$\tau_\mathrm{tc}$ and amplitude~$a_\mathrm{tc}$.
We keep $\tau_\mathrm{tc}$ fixed to 2.5\,ns and use the amplitude~$a_\mathrm{tc}$ to calibrate the sudden phase jump between the interacting $\ket{11}$ and $\ket{20}$ states when the NZTC pulse changes sign~\cite{Negirneac2021}.
We calibrate the total length and the main amplitude of the flux pulse, such that a conditional phase of~$\pi$ and full population recovery from the $\ket{20}$ state is achieved. The average flux pulse length is 71\,ns.
To suppress SWAP errors in the single-excitation manifold, we apply a Gaussian filter of width $\sigma=0.5$\,ns to all flux pulses.
To preclude a possible overlap with preceding and subsequent pulses we add 20\,ns-long buffer periods before and after each flux pulse.
	
Due to the high-pass filtering characteristic of the microwave bias-tee and imperfections in the impedance matching of the flux control line, the flux pulses are subject to distortions on small and long timescales, which we correct for by applying finite impulse response~(FIR) and infinite impulse response~(IIR) filters to the programmed waveforms.
We extract the corresponding IIR filter coefficients from flux pulse scope measurements of the qubit frequency over timescales ranging from 50\,ns to 20\,$\mu$s.
We correct for pulse distortions on the nanosecond timescale by extracting a set of FIR filter coefficients for each qubit using the cryoscope method described in Ref.~\cite{Rol2020}.
To verify accurate pulse control, we measure the time-dependent qubit frequency in response to the NZTC flux pulses, exemplary shown for the gate between Q5 and Q6 in Fig.~\ref{fig:2qubit_gates}(c).	
\\\\	
\textbf{Multiplexed single-shot readout --} To readout the state of all qubits simultaneously, we perform frequency-multiplexed qubit readout as described in Ref.~\cite{Heinsoo2018} using 600\,ns long Gaussian-filtered square pulses. We multiply the digitized readout signal with a set of optimal integration weights, integrate for a period of 650\,ns and  threshold the resulting value to discriminate between the two qubit states.
To evaluate the performance of the qubit readout, we take $n=10$,000 single-shot measurements for each of the 128~possible qubit basis state combinations.
Based on the outcome of an additional preselection readout and for each prepared basis state $i$, we select only those $n_0 \approx 0.91n$ of the shots for further analysis, for which all qubits were initially found in the ground state. For each basis state~$i$, we determine the frequency $f(j|i)$ with which we assigned the state label $j$ to obtain an estimate of the assignment probability matrix $M_{ji} \equiv f(j|i)/n_0$. As shown in Fig.~\ref{fig:ssro}, the assignment error $1-M_{ii}$ increases with excitation number due to qubit relaxation during the readout. As described in the main text, we account for readout errors by multiplying any measured probability distribution $p_i$ with the inverse of $M$ before evaluating expectation values of observables. When qubits are read out individually, e.g. for standard calibration and characterization measurements, we achieve the assignment fidelities stated in Table~\ref{tab:qb_paras}.
\\\\ 
\section{Gate characterization}
\label{GatePerformance}
\begin{figure}[t]
	\centering
	\includegraphics[width = 0.41\textwidth]{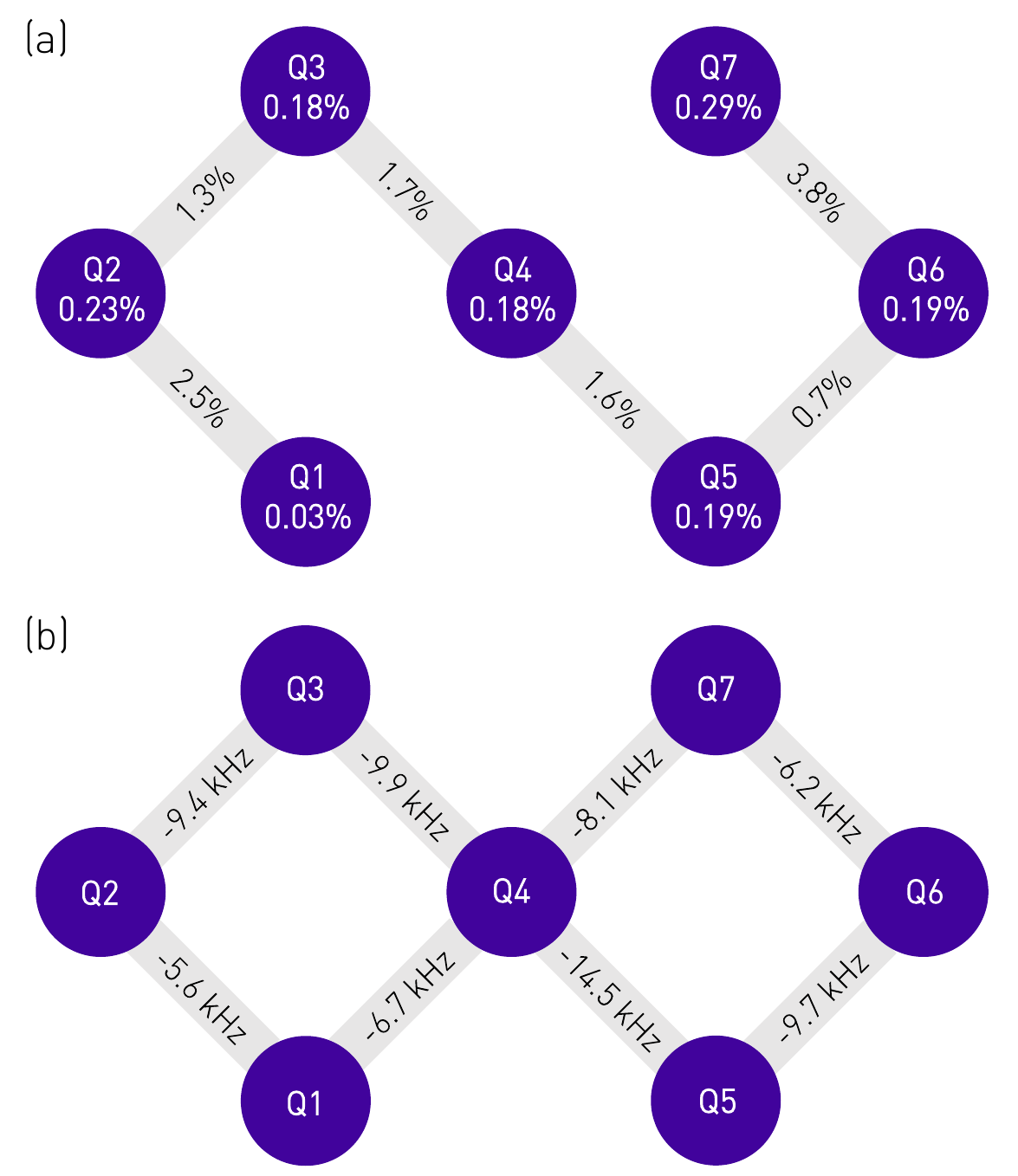}
	\caption{\textbf{Gate performance.} (a) Single and two-qubit gate infidelities, extracted from randomized benchmarking and quantum process tomography, respectively. (b)~Measured residual ZZ coupling strength $\alpha_\mathrm{zz}$ between neighboring qubits at idle frequencies.}
	\label{fig:gates}
\end{figure}
		
\textbf{Single-qubit gate performance --}
We characterize the single-qubit gate performance by individual randomized benchmarking and find an average fidelity of 99.82\,\% per Clifford, see Fig.~\ref{fig:gates}(a).	
To determine the residual ZZ coupling strength $\alpha_\mathrm{zz}$, we measure the shift in qubit frequency using a Ramsey experiment, induced by an excitation of the neighboring qubit~\cite{Collodo2020}. We find values well below 15\,kHz, see Fig.~\ref{fig:gates}(b).
\\\\	
\textbf{Two-qubit gate performance --}		
We calibrate and characterize six CZ gates on the quantum device, such that a qubit connectivity in a 1D chain is established, and use quantum process tomography to evaluate the gate performance by determining the corresponding process $\mathcal{E}(\rho) =\sum_{\alpha,\beta=1}^{16}\chi_{\alpha\beta}E_{\alpha}\rho E_{\beta}^{\dagger}$,  where $E_{\alpha} \in \{I,X,Y,Z \}^{\otimes2}$.
For that purpose, we prepare $R_i|0\rangle$ with $R_i\in \{I, R_x(\pi/2), R_y(\pi/2), R_x(\pi)\}^{\otimes 2}$, apply the CZ gate and perform quantum state tomography for each of the 16 different $R_i$.
To reduce the effect of state preparation errors, we condition the data analysis on having detected the qubits in the ground state initially. To account for readout errors, we multiply the obtained averaged state probabilities $\pmb{p}$ by the inverse of the measured readout assignment probability matrix~$M$ to obtain $\tilde{\pmb{p}}=M^{-1}\pmb{p}$.
Based on the probability distributions $\tilde{\pmb{p}}$, we reconstruct the most likely density matrix $\rho_i$ using a maximum-likelihood procedure. Based on those density matrices, we compute the process matrix $\chi$ following the procedure in Ref.~\cite{Chuang1997}.
We compute the respective gate infidelities $1-F=1-\mathrm{Tr}\left(\chi_\mathrm{cz}\chi\right)$ listed in Fig.~\ref{fig:gates}(a) by comparing $\chi$ to the ideal CZ process matrix $\chi_\mathrm{cz}$ and use the experimentally obtained process matrices for the Kraus operator simulation in Appendix~\ref{Methods}.

\begin{table}[b]
	\begin{tabular}{l c c c c c c c}
		\toprule
		& CZ$_{12}$ & CZ$_{23}$ & CZ$_{34}$ & CZ$_{45}$ & CZ$_{56}$ & CZ$_{67}$  \\ \midrule
		Estimated infidelity (\%)  &2.4&0.5&0.6 &0.6 &0.6 & 2.1\\
		Measured infidelity (\%)  &2.5&1.3&1.7 &1.6 &0.7 & 3.8\\
		\bottomrule
	\end{tabular}
	\caption{Estimated bound for CZ gate infidelities calculated from the measured population loss at the gate interaction frequency
		in comparison to the measured gate infidelities obtained from quantum process tomography.}
	\label{tab:pop_loss}
\end{table}

To determine the contribution of population loss to the two-qubit gate infidelity, we measure the loss at the gate interaction frequency $p_{h,\ket{01}}$ and $p_{l,\ket{10}}$ for both the high and low frequency qubit and the loss $p_{l,\ket{12}}$ at the involved $\ket{12}$ transition frequency, see Fig.\ref{fig:2qubit_gates}(a).
We calculate the expected gate infidelity by modeling the dynamics of our two-qubit gate interaction with a master equation which also accounts for the measured population loss. 
We compute the corresponding process matrix~$\chi$ by evolving the master equation for the duration of the two-qubit gate, determine the expected gate infidelity $1-F=1-\mathrm{Tr}\left(\chi_\mathrm{cz}\chi\right)$ and summarize the obtained values in Tab.~\ref{tab:pop_loss}. We find that the gates CZ$_{12}$ and CZ$_{67}$ are most  strongly affected by population loss during the gate operation, which is consistent with the measured infidelities, see Tab.~\ref{tab:pop_loss}.
\\\\	
\section{Methods}
\label{Methods}
\textbf{Variational circuit optimization --}
	\begin{figure}[t]
		\centering
		\includegraphics[width = 0.49\textwidth]{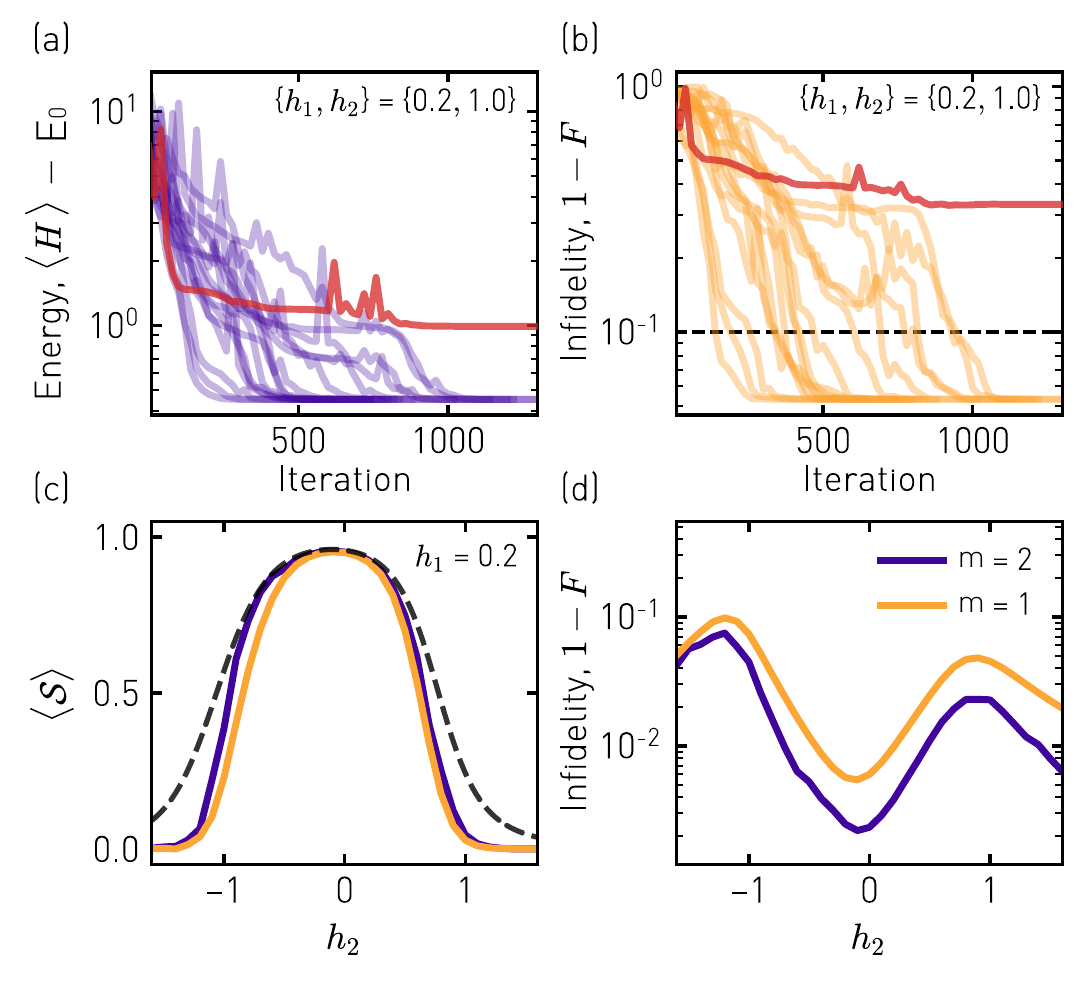}
		\caption{\textbf{Variational circuit optimization.} (a)~Ground state energy $\braket{H}$ as a function of the optimization step for $n=20$ different runs from randomly chosen starting conditions. The energy is stated as a difference to the ideal ground state energy $\braket{H}$ for the specified Hamiltonian parameters~$\{h_1,h_2\}$. (b)~Infidelity of the quantum state as a function of the optimization step with the acceptance threshold of 10\,\% being marked with a dashed black line. The red curves in (a) and (b) depict an unsuccessful optimization run into a local minimum. Value of~$\braket{S}$ in (c) and infidelity $1-F$ in (d) for a fixed $h_1=0.2$ vs. $h_2$ for different variational circuit depths $m$ compared to the ideal value~(dashed line).}
		\label{fig:vqe}
	\end{figure}
As explained in the main text, we simulate the variational state preparation circuit, shown in Fig.~\ref{fig:var_gnd_prep}(a), to minimize the expectation of $\braket{H}$ using the L-BFGS optimizer from the Qiskit Python package~\cite{Abraham2021}. As depicted in Fig.~\ref{fig:vqe}(a) and (b), the circuit optimization typically converges after 500 to 1,000 iterations achieving an average energy accuracy $\braket{H}-E_0\sim0.4$ and average state infidelity $1-F\sim3.5$\,\%, where $E_0$ is the ground state energy obtained from exact diagonalization of $H$. The optimization run marked in red color lies above the acceptance threshold of $1-F>10$\,\% (dashed line in Fig.~\ref{fig:vqe}(b)), which we find to happen more frequently for ground states close to the phase boundary. The challenge of accurately approximating ground state near the phase boundary is also reflected by the larger deviations of $\braket{\mathcal{S}}$ from the ideal value (Fig.~\ref{fig:vqe}(c)) and the larger infidelity (Fig.~\ref{fig:vqe}(d)) in those parameter regions. By increasing the variational circuit depth to $m=2$, the infidelity can be reduced~\cite{BravoPrieto2020}, as depicted in Fig.~\ref{fig:vqe}(d), however under the influence of noise and gate errors, we expect this advantage to diminish and we thus choose to operate our state preparation circuit with $m=1$.
\begin{figure}[t]
	\centering
	\includegraphics[width = 0.48\textwidth]{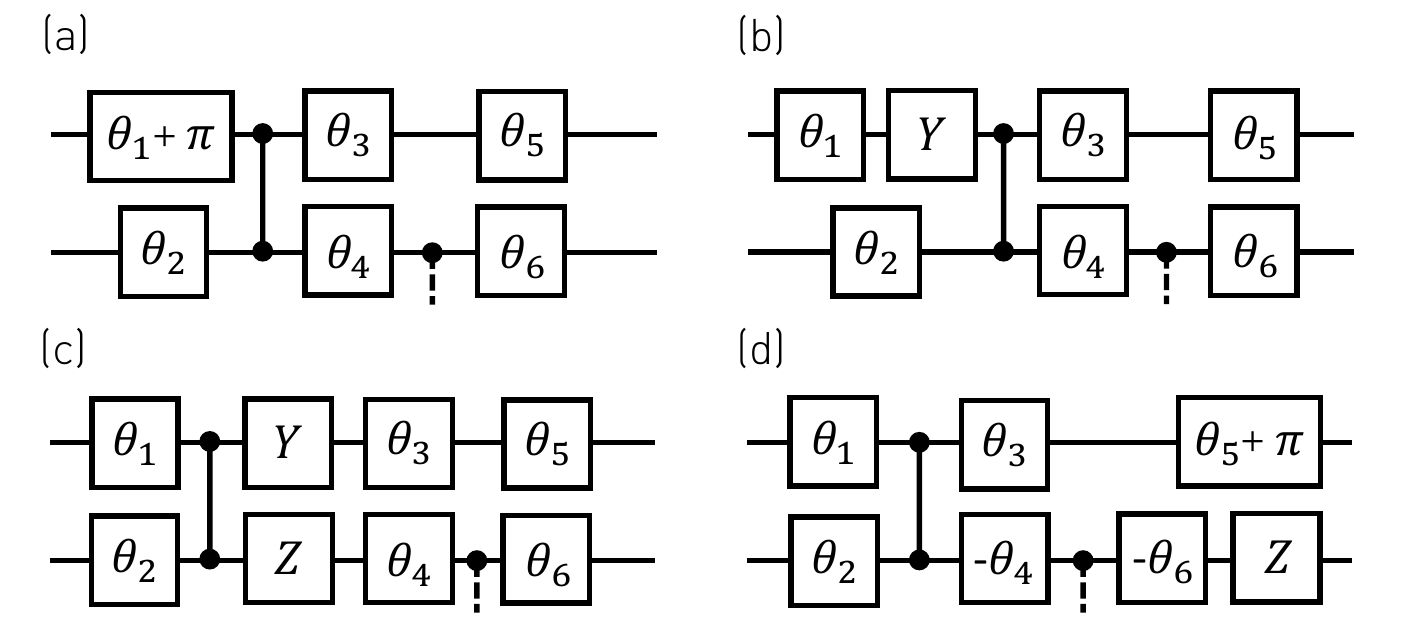}
	\caption{\textbf{Exploiting ansatz symmetries.} Sequence of equivalent quantum circuits showing how to eliminate a rotation angle $\theta_1+ \pi$ close to $\pi$ in the first circuit layer by adjusting the rotation angles in the second and third layer.}
	\label{fig:circ_symm}
\end{figure}
	
\begin{figure}[t]
	\centering
	\includegraphics[width = 0.485\textwidth]{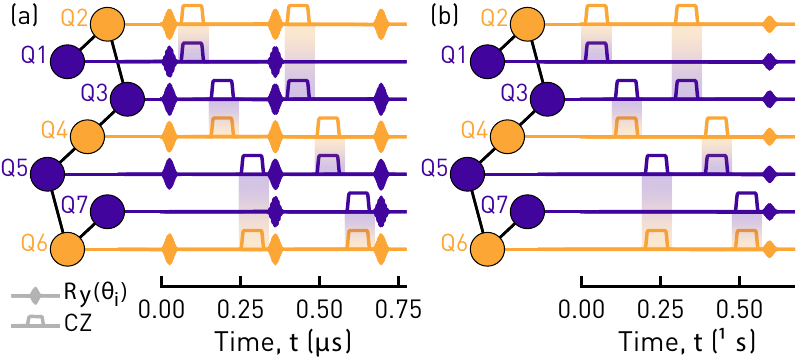}
	\caption{Pulse sequences for implementing (a), the variational ground state preparation and (b), the convolutional layer of the QCNN. The qubit colors represent the different qubit idle frequency bands of $\sim$\,5.9\,GHz (orange) and $\sim$\,4.2\,GHz~(purple).}
	\label{fig:pulse_seq}
\end{figure}

As discussed in the main text, we reduce the effect of qubit relaxation on the state preparation fidelity by finding equivalent parameters $\tilde{\pmb{\theta}}_{\rm opt}$, which result in the same unitary $U({\pmb{\theta}}_{\rm opt})=U(\tilde{\pmb{\theta}}_{\rm opt})$, but which avoid rotation angles $\tilde{{\theta_i}} >\pi/2$ in the first layer of single qubit gates~\cite{Fontana2020a}.
We achieve this by using the identity $R_y(\theta_i+\pi)=R_y(\theta_i)(-iY)$ (see Fig.~\ref{fig:circ_symm}(a) and (b)). The additional $Y$ gate can be propagated through the following CZ gates using the identity  $CZ(Y\otimes I)=(Y\otimes Z)CZ$ (see Fig.~\ref{fig:circ_symm}(c)) and finally be absorbed by the single-qubit rotation in the third layer, yielding the circuit in (d).
We repeat this procedure for all angles $\theta_i$ in the first layer where $|\theta_i| > \pi/2$. This effectively reduces the probability of being in the excited state and thus the rate of amplitude damping on the qubits.
From a Kraus operator simulation of the respective quantum circuits (Appendix~\ref{Theory}), we find that using the equivalent angle set $\tilde{\pmb{\theta}}_{\rm opt}$ can improve the fidelity of the prepared quantum states by up to 10$\,\%$ compared to using~$\pmb{\theta}_{\rm opt}$.
\\\\	
\textbf{Ground state energy --}
\begin{figure}[t]
	\centering
	\includegraphics[width = 0.46\textwidth]{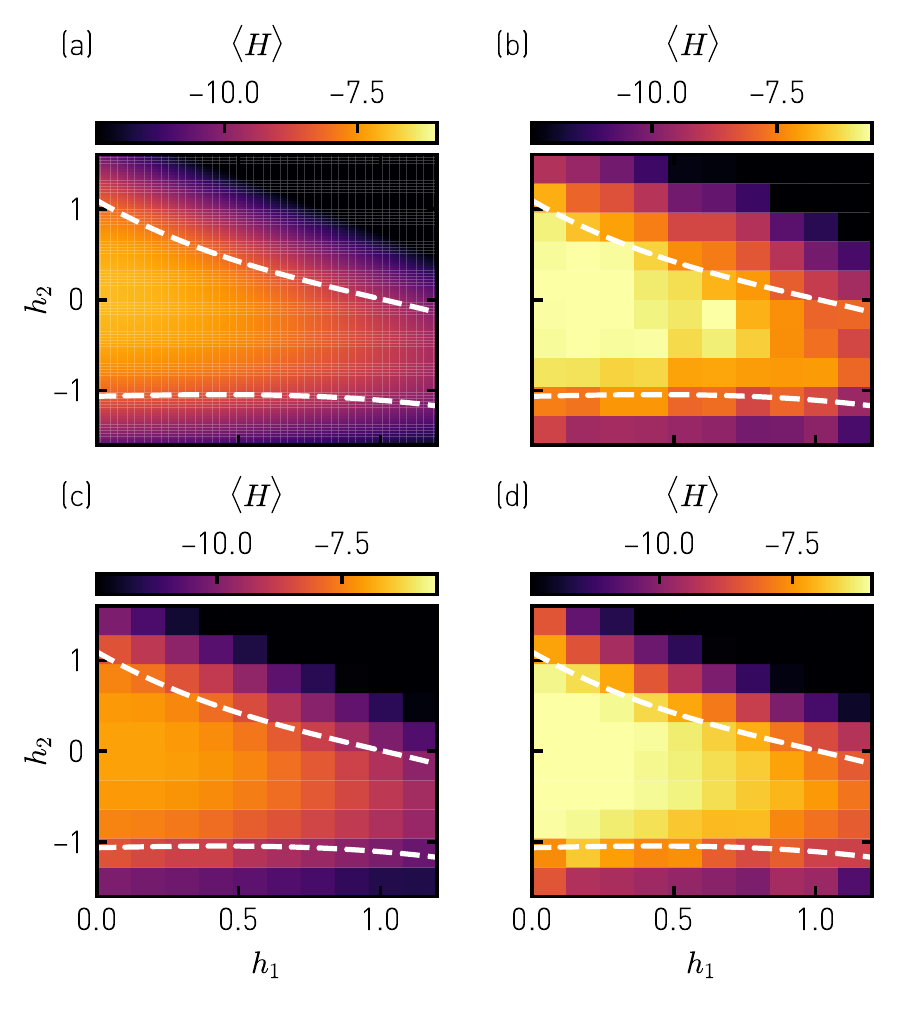}
	\caption{\textbf{Ground state energy.} Energy expectation value $\braket{H}$ with respect to exact ground states in (a), approximate ground states prepared and measured on the quantum hardware in (b), ground states obtained from the ideal state preparation circuit (c) and simulated ground states taking decoherence and gate errors into account in (d).}
	\label{fig:energy}
\end{figure}
As an additional performance measure, complimentary to the fidelity and the string order parameter, we also determine the variational ground state energy in comparison to the exact value, see Fig.~\ref{fig:energy}(a).
We measure $\braket{H}$ by sampling the respective expectation values (see Eq.~\eqref{eq:H}) from the state preparation circuit for 10$\times$10 combinations of $\{h_1,h_2\}$.
We find the measured values to resemble all qualitative features of the exact energy diagram, compare Fig.~\ref{fig:energy}(a) and (b), which additionally confirms the quality of the prepared quantum states also outside the SPT phase where the value of $\braket{\mathcal{S}}$ vanishes.
In addition, we find good agreement between the exact value of the energy and an ideal simulation of the preparation circuit, compare Fig.~\ref{fig:energy}(a) and (c), and most importantly between the measured value of $\braket{H}$ and a Kraus operator simulation taking decoherence and two-qubit gate errors into account, compare Fig.~\ref{fig:energy}(b)~and~(d).
\\\\
\textbf{Kraus operator simulations --} 
To simulate the evolution of a quantum state $\rho$ during the state preparation sequence (Fig.~\ref{fig:pulse_seq}(a)) and the QCNN sequence (Fig.~\ref{fig:pulse_seq}(b)) in the presence of decoherence and gate errors, we model each gate as a completely-positive trace preserving (CPTP) process
	\begin{equation}
		\label{eq: Kraus rep}
		\rho(t+\delta t)=\mathcal{E}(\rho(t)) = \sum_{\alpha=1}^{N}K_{\alpha}\rho(t) K_{\alpha}^{\dagger},
	\end{equation}
where $\delta t$ is the gate duration, $K_{\alpha}$ are Kraus operators satisfying $\sum_{\alpha=1}^{M}K_{\alpha}^{\dagger} K_{\alpha}= I$ and $I$ is the identity matrix. We compose the circuits as sequences of single-qubit, qubit idling, and two-qubit gate processes.

We describe single-qubit $R_y$ gates by the concatenated process $\mathcal{E}_r(\mathcal{E}_{zz}(\mathcal{E}_y(\rho)))$. Here, $\mathcal{E}_y(\rho)=R_y\rho R_y^\dagger$ is the ideal single-qubit unitary, $\mathcal{E}_\mathrm{zz}(\rho) = e^{-i\delta t H_\mathrm{zz}} \rho (e^{-i\delta t H_\mathrm{zz}})^\dagger$ with $H_\mathrm{zz}=\frac{1}{4}\sum_{\langle ij\rangle}\alpha^{ij}_{zz}(I_i-Z_i)(I_j-Z_j)$ describes residual ZZ coupling with strength $\alpha^{ij}_\mathrm{zz}$ of qubit $i$ to its neighbors $j$, and $\mathcal{E}_r(\rho)$ accounts for dephasing and energy relaxation. We model $\mathcal{E}_r$ according to~\eqref{eq: Kraus rep} with Kraus operators $K_1 = \sqrt{\gamma_1}\sigma^{-}$, $K_2 = \sqrt{\gamma_2}Z$ and $K_3 = \sqrt{1-\gamma_2}|0\rangle\langle0|+\sqrt{1-\gamma_1 - \gamma_2}|1\rangle\langle1|$ with $\gamma_1 = \delta t/T_1$ and $\gamma_2 = \delta t(\frac{1}{2T_2} - \frac{1}{4T_1})$. Here, $T_1$, and $T_2$ are taken from experimental characterization experiments, see Tab~\ref{tab:qb_paras}. 

We describe the CZ process directly from experimental data by modeling $\mathcal{E}(\rho) = \sum_{\alpha,\beta=1}^{16}\chi_{\alpha\beta}E_{\alpha}\rho E_{\beta}^{\dagger}$ with $\chi$ being the process matrix obtained from experimental quantum process tomography (see Appendix~\ref{GatePerformance}) and $E_{\alpha} \in \{I,X,Y,Z \}^{\otimes2}$. 
Since CZ gates are executed sequentially, see Fig.~\ref{fig:pulse_seq}, we use the processes $\mathcal{E}_r$ and $\mathcal{E}_{zz}$ on the remaining idle qubits to take into account decoherence and residual ZZ coupling during the gate time.
We implement the Kraus operator simulation of the quantum circuits using the QuTiP Python package~\cite{Johansson2013a}.
\\\\	
\section{Properties of the QCNN circuit}
\label{Theory}
\begin{figure*}[t]
	\centering
	\includegraphics[width = 0.98\textwidth]{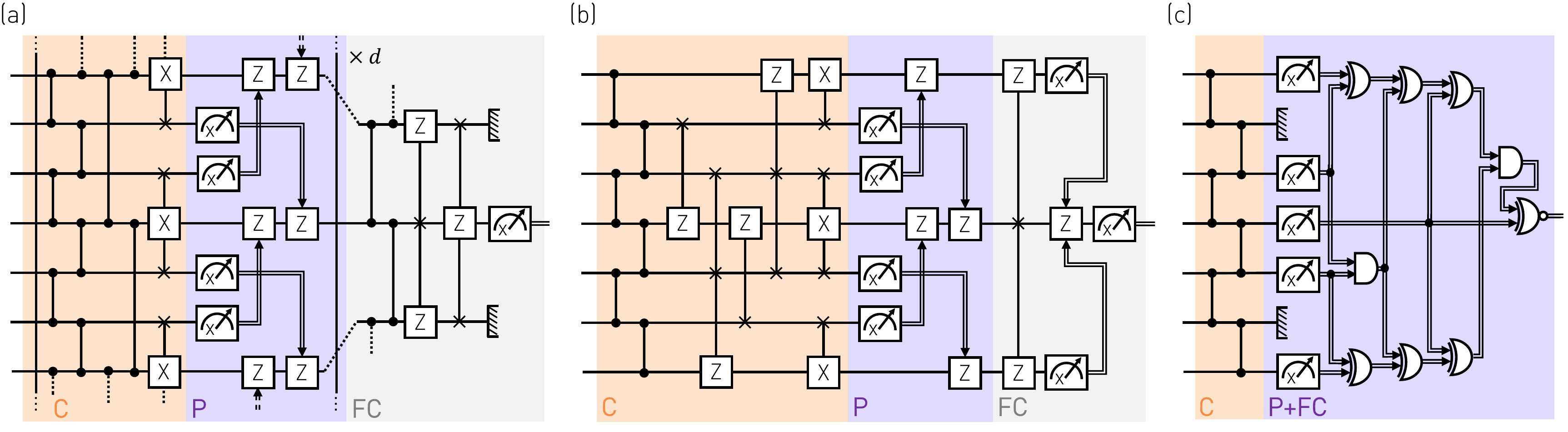}
	\caption{\textbf{QCNN circuit transformation.} (a)~The employed 7-qubit QCNN circuit with quantum gates only~(black) and the extension to a larger system size with the QCNN depth $d$~(dashed), which can be transformed into (b)~an intermediate 7-qubit circuit and into (c)~an equivalent circuit with the pooling and fully connected layer being implemented with classical XOR and AND gates in post-processing.}
	\label{fig:qcnn_circ_mod}
\end{figure*}
\textbf{QCNN circuit design --} The full quantum QCNN circuit as proposed in Ref.~\cite{Cong2019} involves a convolutional (C) layer, a pooling (P) layer and a fully connected (FC) layer, see Fig.~\ref{fig:qcnn_circ_mod}(a). The C layer performs CZ gates with controls in the computational basis as well as a Toffoli (C$_x$C$_x$NOT) gate and two C$_x$NOT gates with control in the $X$ basis, which is triggered when control qubits are in the state~$|-\rangle$. In the P layer, qubits~2, 3, 5 and 6 are measured in the $X$ basis and $Z$ gates are performed on qubits~1, 4 and 7 when measurements yield $X = -1$.

The design of this particular QCNN is inspired by the MERA representation~\cite{Vidal2008} of the cluster state $|\mathcal{C}\rangle$ and it is constructed such that the cluster state becomes a stable fixed point of the QCNN circuit \cite{Cong2019}. The C and P layers transform the cluster state (of 7~qubits) $|\mathcal{C}^{(7)}\rangle \xrightarrow{\rm CP} |\mathcal{C}^{(3)}\rangle$ to the cluster state $|\mathcal{C}^{(3)}\rangle$ of reduced system size (3~qubits). Perturbations of the input state away from the cluster state can be interpreted as "errors". In analogy to quantum error correction, such errors are detected by measurements in the P layer and corrected by conditional $Z$ gates \cite{Cong2019}. In particular, the cluster state perturbed by a single $X_i$ error, for $i=1,2,...,7$, is mapped onto the unperturbed cluster state, $X_i |\mathcal{C}^{(7)}\rangle\xrightarrow{\rm CP}|\mathcal{C}^{(3)}\rangle$. This leads to the convergence of states in the SPT phase towards the cluster state, which enables the QCNN to recognize the phase with high fidelity. 
Such perturbations could either arise in exact ground states due to finite values of $h_1$ and $h_2$ or due to errors in the state preparation. 

While single Pauli X operations are absorbed by the C and P layers, Pauli Z errors propagate through those layers, since $Z_i |\mathcal{C}^{(7)}\rangle\xrightarrow{\rm CP}Z_j|\mathcal{C}^{(3)}\rangle$. To achieve a high fidelity of detecting the SPT phase in the presence of noise, we extent the originally proposed QCNN circuit from Ref.~\cite{Cong2019} and design a new FC layer to correct for $Z$ errors. In particular, this new FC layer involves CZ gates with controls in the computational basis, as well as C$_x$Z gates and a controlled-controlled $Z$ (C$_x$C$_x$Z) gate with controls in the $X$ basis, which are triggered when control qubits are in the state $|-\rangle$, see Fig.~\ref{fig:qcnn_circ_mod}(a). Finally, qubit~4 is measured in the $X$ basis, which together with the preponed FC layer is equivalent to measuring the observables $\frac{1}{2}(Z_1X_4Z_7 + X_1Z_4 + Z_4X_7 + Y_1X_4Y_7)$ on the state after the C and P layers. For the ideal cluster state $|\mathcal{C}^{(3)}\rangle$ as well as for the cluster state $Z_j|\mathcal{C}^{(3)}\rangle$ perturbed by a single $Z_j$ error on qubit $j=1,4$ or 7, this measurement deterministically yields the value $1$ and, as a consequence, it effectively corrects the $Z$ error.

For the paramagnetic and the antiferromagnetic state, the QCNN yields the output zero and thus it can be employed for the recognition of the SPT phase.
\\\\
\textbf{Equivalent QCNN circuit --} Here we derive a circuit that is equivalent to the full quantum QCNN circuit shown in Fig.~\ref{fig:qcnn_circ_mod}(a) but features fewer quantum gates, a projective measurement and classical post-processing, see Fig.~\ref{fig:qcnn_circ_mod}(c). We use a notation, where CZ$_{jk}$ denotes a CZ gate acting symmetrically on qubits $j$ and $k$, and C$_x$NOT$_{j;k}$ denotes a controlled not gate acting on qubit $k$ controlled by qubit $j$ in the $X$ basis (C$_x$C$_x$NOT$_{jl;k}$ denotes a controlled-controlled not gate acting on qubit $k$ controlled by qubits $j$ and $l$ in the $X$ basis).

We first note that $\textrm{CZ}_{14}$ and $\textrm{CZ}_{47}$ gates are performed twice in the full QCNN circuit, see Fig.~\ref{fig:qcnn_circ_mod}(a), namely the first pair in the C layer and the second pair in the FC layer. The second pair commutes with the P layer and thus it can be moved from the FC layer to the C layer. As the CZ gates do not commute with the Toffoli gate $\textrm{C$_x$C$_x$NOT}_{35;4}$, we use the identity $\nobreak{\textrm{CZ}_{47}\textrm{C$_x$C$_x$NOT}_{35;4} = \textrm{C$_x$C$_x$NOT}_{35;4} \textrm{C$_x$C$_x$Z}_{35;7} \textrm{CZ}_{47}}$ and an analogous identity for $ \textrm{CZ}_{14}$ gate, to exchange the second pair of CZ gates and the Toffoli gate by introducing C$_x$C$_x$Z gates with controls on qubits~3 and 5 in the $X$ basis. We use this identity in an analogue fashion to exchange $\textrm{C$_x$NOT}_{2;1}$ and $\textrm{C$_x$NOT}_{6;7}$. Since $\textrm{CZ}^2 = I$, the second pair of CZ gates moved from the FC layer cancels with the first pair in the C layer and we obtain the equivalent circuit shown in Fig.~\ref{fig:qcnn_circ_mod}(b). Note that we also replaced the last gate of the FC layer, the C$_x$C$_x$Z gate, by equivalent measurements of qubits~1 and~7 in the $X$ basis and the $Z_4$ gate conditioned on measuring $X_1=X_7=-1$.

We will now show that in the equivalent quantum circuit depicted in Fig.~\ref{fig:qcnn_circ_mod}(c), all gates except from CZ gates with controls in the computational basis can be performed in classical post-processing after the projective measurement of all qubits in the $X$ basis. The Toffoli gate is directly followed by the measurement of the control qubits~3 and~5 and it anticommutes with the remaining $Z_4$ gates before the measurement of qubit~4. As a result, the Toffoli gate can produce only an undetectable global phase factor $-1$ and thus it can be removed. $Z$ gates produce a phase flip $|\pm\rangle\rightarrow|\mp\rangle$, which can be implemented in classical post-processing after the measurement in the $X$ basis with eigenstates~$|\pm\rangle$. Conditional $Z$ gates correspond to XOR gates in post-processing. Controlled $Z$ gates with controls in the $X$ basis, which are directly followed by the measurement of control qubits, are equivalent to conditional $Z$ gates and thus they can  also be analogously implemented in post-processing. Moving all conditional $Z$ gates and controlled $Z$ gates with controls in the $X$ basis, one by one, into post-processing, we obtain the equivalent circuit shown in Fig.~\ref{fig:qcnn_circ_mod}(c). This circuit is equivalent to the full quantum QCNN circuit but involves only two layers of CZ gates, projective measurement and classical post-processing.
\\\\
\textbf{Multi-scale string order parameter --} The observable measured at the output of the FC layer corresponds to measuring a multi-scale string order parameter of the form 
\begin{align}
	\mathcal{S}_{\rm M} =\sum_{jk}\eta^{(1)}_{jk}\mathcal{S}_{jk}+\sum_{jklm}\eta^{(2)}_{jklm}\mathcal{S}_{jk}\mathcal{S}_{lm} + ...,\label{eq:MSOP}
\end{align}
where
\begin{align}
	\mathcal{S}_{jk}=Z_jX_{j+1}X_{j+3}...X_{k-3}X_{k-1}Z_k,
\end{align}
are string order parameters of varying length and $\eta^{(\alpha)}$ are coefficients weighting the individual terms~\cite{Cong2019}. 
We now determine the multi-scale SOP measured by the QCNN implemented in the experiment. We choose the termination at the edges of the circuit as shown in Fig.~\ref{fig:qcnn_circ_mod}(a). For $d$ layers of convolution and pooling this choice corresponds to $N=2\cdot 3^{d+1}-11$ qubits at the input of the QCNN.
For the analysis of the full QCNN circuit, depicted in Fig.~\ref{fig:qcnn_circ_mod}(a), we replace measurements and conditional unitaries in the P layers by controlled Z gates with controls in the $X$ basis. In this way, we can represent the QCNN circuit by a unitary $U=U_{\rm FC} V$, where $V=U^{(d)}_{\rm CP}...U^{(1)}_{\rm CP}$ and $U^{(f)}_{\rm CP} = U^{(f)}_{\rm P}U^{(f)}_{\rm C}$, consisting of $d$ convolutional $U^{(f)}_{\rm C}$ and pooling $U^{(f)}_{\rm P}$ layers, $f=1,2,\dots,d$, as well as the fully connected layer $U_{\rm FC}$, which represent the unitary operations of the respective circuit layers shown in Fig.~\ref{fig:qcnn_circ_mod}(a).

The outcome of measuring the output qubit with index $i=\frac{N+1}{2}$ in the FC layer of the QCNN corresponds to the expectation value of the multi-scale~SOP
\begin{align}
	\mathcal{S}_{\rm M} =& V^{\dagger} U_{\rm FC}^{\dagger}X_{\frac{N+1}{2}}U_{\rm FC} V = \frac{1}{2}\left(\mathcal{S}_{\rm M}^{I} - \mathcal{S}_{\rm M}^{II}\right),\label{eq:MSOP1}
\end{align}
where 
\begin{align}
	\mathcal{S}_{\rm M}^{I} &=V^{\dagger}\left(C^{(d)}_{\frac{N+1}{2}-3^d} +C^{(d)}_{\frac{N+1}{2}} +C^{(d)}_{\frac{N+1}{2}+3^d}\right)V,\label{eq:MSOPI}\\
	\mathcal{S}_{\rm M}^{II} &=  V^{\dagger}C^{(d)}_{\frac{N+1}{2}-3^d}C^{(d)}_{\frac{N+1}{2}}C^{(d)}_{\frac{N+1}{2}+3^d}V,\label{eq:MSOPII}
\end{align} 
with $C^{(f)}_j = Z_{j-3^f}X_jZ_{j+3^f}$. To explicitly evaluate the multi-scale SOP, we use the recursive relation
\begin{align}
	U^{(f)\,\dagger}_{\rm CP}&\left(C^{(f)}_{j}C^{(f)}_{j+2 \cdot3^f}...C^{(f)}_{k}\right)U^{(f)}_{\rm CP}\nonumber\\
	&=L^{(f-1)}_{j}\left(C^{(f-1)}_{j}C^{(f-1)}_{j+2\cdot3^{f-1}}...C^{(f-1)}_{k}\right)R^{(f-1)}_{k},\label{eq:pol}
\end{align}
where $k\geq j$ and
\begin{align}
L^{(f)}_{{j}}=&\frac{1}{2}\left(C^{(f)}_{{j}-4 \cdot3^f}C^{(f)}_{{j}-2 \cdot3^f}-C^{(f)}_{{j}-4 \cdot3^f}+C^{(f)}_{{j}-2 \cdot3^f}+1\right),\\
R^{(f)}_{{j}}=&\frac{1}{2}\left(1 + C^{(f)}_{{j}+2 \cdot3^f} - C^{(f)}_{{j}+4 \cdot3^f} + C^{(f)}_{{j}+2 \cdot3^f}C^{(f)}_{{j}+4 \cdot3^f}\right).
\end{align}
Terms at zero depth, i.e. measured at the input state, and at the edge of the array reduce to $L^{(0)}_1=L^{(0)}_4=R^{(0)}_{N-3}=R^{(0)}_N=1$ and we use the notation $X_j=Z_j=\mathbb{1}$ for $j>N$ and $j<1$. Using the recursive relation \eqref{eq:pol} $d$ times, we obtain the explicit form \eqref{eq:MSOP} of the multi-scale SOP $\mathcal{S}_M$ from Eq.~\eqref{eq:MSOP1}. 
To evaluate the second part $\mathcal{S}_{\rm M}^{II}$ of the multi-scale SOP, we exploit that individual $C_i^{(d)}$ terms commute to express $\mathcal{S}_{\rm M}^{II}= V^{\dagger}C^{(d)}_\frac{N+1}{2}VV^{\dagger}C^{(d)}_{\frac{N+1}{2}-3^d}C^{(d)}_{\frac{N+1}{2}+3^d}V$ and use the recursive relation \eqref{eq:pol} separately for $C^{(d)}_{\frac{N+1}{2}}$ and~$C^{(d)}_{\frac{N+1}{2}-3^d}C^{(d)}_{\frac{N+1}{2}+3^d}$. 
\\\\
For the QCNN with the depth $d$, the multi-scale SOP involves $\mathcal{O}(256^{3^{d-1}})$ products of $\mathcal{O}(3^{2d})$ different SOPs. The maximal length of the SOPs at depth $d$ scales as~$\mathcal{O}(3^d)$.

As an example case, the output of the QCNN implemented in the experiment with depth $d=1$ and $N=7$ qubits corresponds to the expectation value of the multi-scale SOP
\begin{align}
	\mathcal{S}_{\rm M} =& U_{\rm CP}^{\dagger} U_{\rm FC}^{\dagger}X_4U_{\rm FC} U_{\rm CP}\\
	=&\frac{1}{2}U_{\rm CP}^{\dagger} (C^{(1)}_1 + C^{(1)}_4 + C^{(1)}_7 - C^{(1)}_1C^{(1)}_4C^{(1)}_7)U_{\rm CP}\\
	=&\sum_{jk}\eta^{(1)}_{jk}\mathcal{S}_{jk}+\sum_{jklm}\eta^{(2)}_{jklm}\mathcal{S}_{jk}\mathcal{S}_{lm}\label{eq:MSOP7}
\end{align}	
By evaluating the unitary transformation $U_{\rm CP}$, we obtain the explicit form  \eqref{eq:MSOP7} of the multi-scale SOP with products of SOPs $\mathcal{S}_{jk}$ weighted with coefficients $\eta^{(\alpha)}$ which are listed in Tab~\ref{tab:SM}. Due to the small system size and the shallow depth $d=1$, the total number of terms measured by the QCNN reduces to 10. Also the maximal length of the SOPs is limited by the finite system size. 
\begin{table}[t]
	\begin{tabular}{ c c l c c c c c c c c }
		\toprule 
		$\eta^{(\alpha)}$ & & Product   &    & \multicolumn{7}{l}{Pauli string}  \\
		\midrule 
		$+1/4$& & $\mathcal{S}_{02}$ & & $X_1$ & $Z_2$ &       &       &       &       &       \\
		$+1/4$& & $\mathcal{S}_{04}$ & & $X_1$ &       & $X_3$ & $Z_4$ &       &       &       \\
		$+1/4$& & $\mathcal{S}_{06}$ & & $X_1$ &       & $X_3$ &       & $X_5$ & $Z_6$ &       \\
		$+1/4$& & $\mathcal{S}_{28}$ & &       & $Z_2$ & $X_3$ &       & $X_5$ &       & $X_7$ \\
		$+1/4$& & $\mathcal{S}_{48}$ & &       &       &       & $Z_4$ & $X_5$ &       & $X_7$ \\
		$+1/4$& & $\mathcal{S}_{68}$ & &       &       &       &       &       & $Z_6$ & $X_7$ \\
		$-1/4$& & $\mathcal{S}_{02}\mathcal{S}_{46}$ & & $X_1$ & $Z_2$ &       & $Z_4$ & $X_5$ & $Z_6$ &       \\
		$-1/4$& & $\mathcal{S}_{24}\mathcal{S}_{68}$ & &       & $Z_2$ & $X_3$ & $Z_4$ &       & $Z_6$ & $X_7$ \\
		\midrule 
		$+1/2$& &$\mathcal{S}_{35}$  & &     &     & $Z_3$ & $X_4$ & $Z_5$ &     &     \\
		\midrule 
		$-1/2$& &$\mathcal{S}_{08}\mathcal{S}_{35}$  & & $X_1$ &     & $Y_3$ & $X_4$ & $Y_5$ &     & $X_7$ \\
		\bottomrule 
	\end{tabular}
	\caption{Individual products of SOPs involved in the multi-scale string order parameter $\mathcal{S}_M$ \eqref{eq:MSOP7} measured by the employed 7-qubit QCNN circuit from Fig.~\ref{fig:qcnn_circ_mod}(a) with $d=1$. The terms are sorted and subdivided according to their measurement basis.} 
	\label{tab:SM}
\end{table}
\\\\
\textbf{Direct measurement of QCNN output --} Instead of performing the QCNN, we could determine the expectation value of the multi-scale SOP $\mathcal{S}_{\rm M}$ \eqref{eq:MSOP} by directly measuring the individual products of SOPs on the input state to obtain the same outcome. We now discuss a crucial reduction in the number of measurements provided by the QCNN compared to the direct sampling from the input state. This in turn leads to the reduction of the sampling complexity --- the number of input-state copies required for determining the expectation value of the multi-scale SOP $\mathcal{S}_{\rm M}$ --- since for each projective measurement we need to prepare a separate copy of the input state.

The multi-scale SOP $\mathcal{S}_{\rm M}$  involves $\mathcal{O}(256^{3^{d-1}})$ products of SOPs, see Eq.~\eqref{eq:MSOP}. A single projective measurement of the input state consists of measuring all qubits in a local basis ($X$, $Y$ or $Z$ basis). We can use classical post-processing to determine the expectation value of several products of SOPs from the same measurement. In particular, two products of SOPs can be sampled from the same measurement if they involve the same Pauli operator on all qubits, on which both products of SOPs act non-trivially. A product of SOPs acts trivially (non-trivially) on qubit $i$ if it involves the identity $I_i$ (the Pauli $X_i$, $Y_i$ or $Z_i$ operator).

The QCNN proposed in Ref.~\cite{Cong2019} measures only the first part $\mathcal{S}_{\rm M}^{I}$ of the multi-scale SOP $\mathcal{S}_{\rm M}$, see Eq.~\eqref{eq:MSOPI}. It can be shown that most of the products of SOPs involved in $\mathcal{S}_{\rm M}^{I}$ can be sampled from the same measurement. The recursive relation \eqref{eq:pol} dictates that, for an arbitrary depth $d$ of the QCNN, all products of SOPs involved in $\mathcal{S}_{\rm M}^{I}$ exhibit either (i) the Pauli $X$ operator at even qubits and Pauli $Z$ operator or identity at odd qubits, or (ii) the Pauli $X$ operator at odd qubits and Pauli $Z$ operator or identity at even qubits. As a result, using classical post-processing, we can determine the expectation values of all products of SOPs involved in $\mathcal{S}_{\rm M}^{I}$ from only two different measurements, namely (i) the measurement of all even qubits in the $X$ basis and all odd qubits in the $Z$ basis, and  (ii) the measurement of all odd qubits in the $X$ basis and all even qubits in the $Z$ basis. This shows that the output of the QCNN proposed in Ref.~\cite{Cong2019} can be efficiently sampled directly from the input state. The direct sampling requires only twice as many measurements as sampling after performing that QCNN with an arbitrary depth.

The number of direct measurements largely increases due to the modification of the FC layer that we consider in this work. Due to the modification, the QCNN measures also the second part $\mathcal{S}_{\rm M}^{II}$ of the multi-scale SOP $\mathcal{S}_{\rm M}$, see Eq.~\eqref{eq:MSOPII}. Using the recursive relation \eqref{eq:pol}, we find out that the products of SOPs involved in $\mathcal{S}_{\rm M}^{II}$ do not follow the simple pattern of $X$ operators at even/odd qubits and $Z$ operators or identities at odd/even qubits. 
It can be shown that, for the QCNN of depth $d>1$, at least $16^{3^{d-1}}$ products of SOPs involved in $\mathcal{S}_{\rm M}^{II}$ require a separate measurement with readout of qubits in different bases. 

As a result, the number of direct measurements on the input state increases double-exponentially with the depth of the QCNN, which renders the direct sampling of the QCNN output unfeasible. On the other hand, performing the QCNN enables us to efficiently sample the multi-scale SOP $\mathcal{S}_{\rm M}$ from a single measurement of the single qubit at the output of the QCNN.

\bibliography{qcnnRefDB}
	
\end{document}